\documentclass[12pt]{article}
\pdfoutput=1
\usepackage[height=8.85in,width=6.45in]{geometry}

\usepackage[utf8]{inputenc}
\usepackage{amsmath,amssymb,amsthm,mathtools}
\usepackage{times,courier,bm,booktabs,array,graphicx}
\usepackage[svgnames,dvipsnames]{xcolor}
\usepackage[colorlinks,citecolor=DarkGreen,linkcolor=FireBrick,urlcolor=FireBrick,linktocpage,unicode,psdextra]{hyperref}
\usepackage{cite}
\usepackage{dsfont}
\usepackage[bottom]{footmisc}
\usepackage{tikz}
\usetikzlibrary{arrows.meta,positioning,calc,decorations.markings}
\numberwithin{equation}{section}
\newcommand{\bea}{\begin{eqnarray}}
\newcommand{\eea}{\end{eqnarray}}
\def\no{\nonumber}
\def\ZZ{\mathbb{Z}}

\def\cA{\mathcal{A}}
\def\cB{\mathcal{B}}

\def\cH{\mathcal{H}}

\def\cZ{\mathcal{Z}}
\newcommand{\Tr}{\mathrm{Tr}}

\newtheorem{proposition}{Proposition}[section]

\theoremstyle{remark}

\begin{document}
\begin{titlepage}
\begin{flushright}
KYUSHU-HET-380
\end{flushright}
\vskip 2.5cm
\begin{center}
{\Large\bfseries Hecke Relations from Self-Duality}
\vskip 1.2cm
Justin Kaidi
\vskip .8cm
 Department of Physics, Kyushu University, Fukuoka 819-0395, Japan\\
 Institute for Advanced Study, Kyushu University, Fukuoka 819-0395, Japan\\
 Quantum and Spacetime Research Institute (QuaSR), Kyushu University, Fukuoka 819-0395, Japan

\end{center}
\vskip 1.2cm
\noindent
\begin{quote}
Hecke operators can be used to relate characters of rational conformal field theories with different central charges, but it has remained a mystery how to interpret these relations physically. We offer a simple explanation in terms of self-duality under asymmetric orbifolds. Self-dual orbifolds also give more general identities beyond the familiar Harvey-Wu construction, which we illustrate through examples.
\end{quote}
\end{titlepage}
\setcounter{tocdepth}{2}

\tableofcontents

\section{Introduction}

Hecke operators can be used to relate the characters of rational conformal field theories with different central charges. Harvey and Wu constructed such operators for vector-valued modular forms and found a number of striking examples \cite{Harvey:2018rdc}. These include relations between $A_{1,1}$ and $E_{7,1}$, between the Lee-Yang model and $G_{2,1}$ or $F_{4,1}$, and between Ising and several free-fermion and exceptional theories. Further work clarified the action of Hecke operators on modular data and extended the construction to additional character relations \cite{Harvey:2019qzs,Duan:2022ltz}. Related identities between ${\mathfrak{sl}}_2$ characters and Virasoro minimal-model characters were derived in \cite{Adamovic:2025efe}. Our goal here is to understand the physical origin of some of these Hecke relations.

The rescaled modular parameters in a Hecke sum suggest a connection to permutation orbifolds. For prime $p$, the connected degree-$p$ covers of a torus have modular parameters $p\tau$ and $(\tau+j)/p$, with $j=0,\ldots,p-1$. As we review in Section \ref{sec:PermutationCovers}, precisely these arguments occur when a symmetry cyclically permutes $p$ identical CFT factors: following a field through the twisted boundary condition gives a single copy of the original theory on a covering surface \cite{Bantay:1997ek,Dijkgraaf:1996xw}. The remaining question is why the sum of these contributions should give the characters of some other CFT.

In this work, we will explain a broad class of examples using asymmetric orbifolds \cite{Narain:1986qm,Narain:1990mw}. By asymmetric, we mean that the symmetry acts on the left-movers and leaves the right-movers unchanged; this allows us to keep the right-moving characters as labels and read off separate identities for the left-moving characters. 
Let $\cB$ denote a theory, and let $g$ generate an anomaly-free $\ZZ_p$ symmetry. Write $Z^s_r$ for the torus trace over the $g^r$-twisted Hilbert space with insertion of $g^s$. Gauging gives
\bea \label{eq:introorbifold}
Z_{\cB/\ZZ_p}={1\over p}\sum_{r,s=0}^{p-1}Z^s_r~.
\eea
If the orbifold returns $\cB$, or more generally has the same torus partition function, we can move the contribution $Z^0_0=Z_{\cB}$ to the left-hand side to obtain
\bea \label{eq:introselfduality}
Z_{\cB}={1\over p-1}\sum_{(r,s)\ne(0,0)}Z^s_r~.
\eea
The nonzero backgrounds fall into $p+1$ classes of $p-1$ elements. Now suppose that the trace with a symmetry insertion can be expressed through characters of another theory $\cA$ at $p\tau$, which is the modular parameter of one of the $p$-fold covering tori. Modular transformations then give the other covering tori, and (\ref{eq:introselfduality}) gives a Hecke relation.\footnote{The orbifold sum (\ref{eq:introorbifold}) and its rearrangement (\ref{eq:introselfduality}) apply to both symmetric and asymmetric orbifolds. It is the extraction of individual character identities, with the right-moving sectors held fixed, that will use the asymmetric action.} 

 A literal permutation is one way to obtain the required symmetry-inserted traces. We will also encounter other internal symmetries, such as rotations of currents and finite symmetries of commutant theories. Their traces have the same dependence on the covering parameter $p\tau$, even though the symmetry does not permute copies of $\cA$.  The general framework will be given in Section \ref{sec:criterion}, but it is useful to begin here with a concrete example. Throughout, $\chi^{\cA}_0$ denotes the vacuum character of $\cA$. Other subscripts specify either the current-algebra representation or its lowest conformal weight, as stated in each example. 

\subsection{A first example: $A_{1,1}$ and $E_{7,1}$}

To illustrate the basic idea, let us begin by understanding the Hecke identity relating $A_{1,1}$ characters to those of $E_{7,1}$. 
The two theories have central charges $1$ and $7$. Their two irreducible modules have lowest conformal weights $(0,1/4)$ and $(0,3/4)$, respectively. We write their characters explicitly as
\bea
\chi^{A_{1,1}}_0(\tau)&=&q^{-1/24}(1+3q+4q^2 + 7 q^3 + \cdots)~,\no\\
\chi^{A_{1,1}}_{1/4}(\tau)&=&q^{5/24}(2+2q+6 q^2 + 8 q^3 + \cdots)~,\no\\
\chi^{E_{7,1}}_0(\tau)&=&q^{-7/24}(1+133q+1673 q^2 + 11914 q^3 + \cdots)~,\no\\
\chi^{E_{7,1}}_{3/4}(\tau)&=&q^{11/24}(56+968 q + 7504 q^2 + 42616q^3 + \cdots)~,
\eea
which we organize into vectors $\boldsymbol{\chi}^{A_{1,1}} = (\chi^{A_{1,1}}_0, \chi^{A_{1,1}}_{1/4})$ and $\boldsymbol{\chi}^{E_{7,1}} = (\chi^{E_{7,1}}_0, \chi^{E_{7,1}}_{3/4})$. 
Both character vectors have the same modular $S$ matrix,
\bea
\label{eq:A11Smatrix}
S_{A_{1,1}} = S_{E_{7,1}}= 2^{-1/2}\left(\begin{matrix}1&1\\1&-1\end{matrix}\right)~.
 \eea

In \cite{Harvey:2018rdc}, a relation is given which expresses the $E_{7,1}$ characters as sums of the corresponding $A_{1,1}$ characters on eight degree-seven covering tori. 
In terms of the Harvey-Wu operators $\mathsf{T}_p$ defined below in (\ref{eq:ordinaryhecke}), this relation reads
\bea \label{eq:HarveyWuexample}
\boldsymbol{\chi}^{E_{7,1}} = \mathsf{T}_7\,  \boldsymbol{\chi}^{A_{1,1}}~.
\eea
We will now derive this expression from self-duality under a certain asymmetric orbifold of the diagonal $E_{7,1}$ theory.

To begin, consider the conformal embedding $A_{1,1}^{\otimes7}\subset E_{7,1}$ \cite{Schellekens:1986mb}. Let $V_{A_1}$ and $M_{A_1}$ denote the vacuum and weight-$1/4$ modules of $A_{1,1}$, and let $g$ be the symmetry which cyclically permutes the seven factors. The $E_{7,1}$ vacuum decomposes into the following eight modules,
\bea \label{eq:EsevenModuleBranching}
V_{E_7}=V_{A_1}^{\otimes7}\oplus\bigoplus_{j=0}^6 g^j\!\left(M_{A_1}^{\otimes3}\otimes V_{A_1}\otimes M_{A_1}\otimes V_{A_1}^{\otimes2}\right)~.
\eea
In each of the seven additional summands, four factors are in $M_{A_1}$ and three are in $V_{A_1}$. These summands have lowest weight one and contain the currents that extend $A_1^7$ to $E_7$. The non-vacuum $E_{7,1}$ module is obtained by interchanging $V_{A_1}$ and $M_{A_1}$ in every factor of this decomposition. It therefore contains $M_{A_1}^{\otimes7}$ and seven summands with three factors in $M_{A_1}$. Thus the characters decompose as
\bea \label{eq:EsevenBranching}
\chi^{E_{7,1}}_0&=&(\chi^{A_{1,1}}_0)^7+7(\chi^{A_{1,1}}_0)^3(\chi^{A_{1,1}}_{1/4})^4~,
\no\\
\chi^{E_{7,1}}_{3/4}&=&(\chi^{A_{1,1}}_{1/4})^7+7(\chi^{A_{1,1}}_0)^4(\chi^{A_{1,1}}_{1/4})^3~.
\eea
All characters in this equation have argument $\tau$. 

Inserting any non-identity power $g^s$ into the trace gives zero on the seven summands that are cyclically exchanged, leaving only $V_{A_1}^{\otimes7}$. In the other module, only $M_{A_1}^{\otimes7}$ contributes. As we review in Section \ref{sec:PermutationCovers}, the trace of a seven-cycle on seven identical factors is the trace of one factor with argument $7\tau$. Hence
\bea \label{eq:EsevenTwining}
\Tr_{V_{E_7}}g^s q^{L_0-7/24}&=&\chi^{A_{1,1}}_0(7\tau)~,\no\\
\Tr_{M_{E_7}}g^s q^{L_0-7/24}&=&\chi^{A_{1,1}}_{1/4}(7\tau)~,
\hspace{0.4 in}s=1,\ldots,6~,
\eea
where $M_{E_7}$ is the weight-$3/4$ module of $E_{7,1}$.  The symmetry-inserted traces of $E_{7,1}$ can thus be written in terms of the characters of $A_{1,1}$  on the seven-fold covering torus with modular parameter $7\tau$. 

Furthermore, the diagonal $E_{7,1}$ CFT can be shown to be self-dual under the asymmetric orbifold by this symmetry. A simple way to understand this is the following.  The VOA $E_{7,1}$ admits a holomorphic completion
\bea \label{eq:EsevenSpectatorBranching}
V_{E_8} =(V_{E_7}\otimes V_{A_1})\oplus(M_{E_7}\otimes M_{A_1})~,
\eea
where the seven-cycle extends to an automorphism fixing the additional $A_{1,1}$ factor. Its twisted sector weight is $2/7$, as derived in Appendix \ref{app:E7}. The condition $7h_g=2\in\ZZ$ ensures that the symmetry is anomaly-free. Gauging the symmetry then gives another holomorphic theory of central charge eight, which must again be $E_{8,1}$. The spectator $A_{1,1}$ survives with its original level. A level-one $A_1$ inside $E_{8,1}$ is conjugate to a root $A_1$, and its commutant is $E_{7,1}$. 
Pairing this with the unchanged right-moving modules then gives the original diagonal $E_{7,1}$ CFT. 

The upshot is that we now have both of the ingredients---a symmetry-inserted trace on a covering torus, and a self-duality under gauging---needed to obtain a Hecke relation. To obtain the relation, we compute the explicit twisted-twined partition functions. Suppressing the common argument $\tau$ of the right-moving characters, we have
\bea
Z^0_0&=&|\chi^{E_{7,1}}_0|^2+|\chi^{E_{7,1}}_{3/4}|^2~,
\no\\
Z^s_0&=&\chi^{A_{1,1}}_0(7\tau)\overline{\chi^{E_{7,1}}_0}
+\chi^{A_{1,1}}_{1/4}(7\tau)\overline{\chi^{E_{7,1}}_{3/4}}~, \hspace{0.55 in}s\ne0~,
\no\\
Z^0_r&=&\chi^{A_{1,1}}_0(\tau/7)\overline{\chi^{E_{7,1}}_0}+\chi^{A_{1,1}}_{1/4}(\tau/7)\overline{\chi^{E_{7,1}}_{3/4}}~,
\hspace{0.4 in}r\ne0~,
\eea
with the last line following from (\ref{eq:A11Smatrix}). 
The other backgrounds follow from $Z^s_r(\tau+1)=Z^{s+r}_r(\tau)$. For $r\ne0$ and $s=rj\pmod7$, this gives
\bea \label{eq:EsevenAllTraces}
Z^{rj}_r= e^{2\pi i\,7j/24}\chi^{A_{1,1}}_0\!\left({\tau+j\over 7}\right)
\overline{\chi^{E_{7,1}}_0} +e^{-2\pi i\,11j/24}\chi^{A_{1,1}}_{1/4}\!\left({\tau+j\over 7}\right)
\overline{\chi^{E_{7,1}}_{3/4}}~.
\eea
The phases come from the right-moving $E_{7,1}$ characters. 

Since the order of the $T$-matrix of $A_{1,1}$ is $24$, and since multiplication by $24$ permutes the residues modulo seven, the sums over $j$ can be written without explicit phases as,
\bea
\sum_{j=0}^6e^{2\pi i\,7j/24}\chi^{A_{1,1}}_0\!\left({\tau+j\over 7}\right)
&=&\sum_{n=0}^6\chi^{A_{1,1}}_0\!\left({\tau+24n\over 7}\right)~,\no\\
\sum_{j=0}^6e^{-2\pi i\,11j/24}\chi^{A_{1,1}}_{1/4}\!\left({\tau+j\over 7}\right)
&=&\sum_{n=0}^6\chi^{A_{1,1}}_{1/4}\!\left({\tau+24n\over 7}\right)~.
\eea
There are six equal contributions from insertions of $g^s$, and six equal sums from the  $g^r$-twisted sectors. The coefficient of $\overline{\chi^{E_{7,1}}_0}$ in the orbifold partition function is consequently
\bea
{1\over 7}\left[
\chi^{E_{7,1}}_0(\tau)+6\chi^{A_{1,1}}_0(7\tau)
+6\sum_{n=0}^6\chi^{A_{1,1}}_0\!\left({\tau+24n\over 7}\right)
\right]~.
\eea
Self-duality says that this coefficient is equal to $\chi^{E_{7,1}}_0(\tau)$. Applying the same argument to the other right-moving character yields
\bea
\label{eq:E7A1finalrelation}
\chi^{E_{7,1}}_0(\tau)
&=&\chi^{A_{1,1}}_0(7\tau)
+\sum_{n=0}^{6}\chi^{A_{1,1}}_0\!\left({\tau+24n\over 7}\right)~,\no\\
\chi^{E_{7,1}}_{3/4}(\tau)
&=&\chi^{A_{1,1}}_{1/4}(7\tau)
+\sum_{n=0}^{6}\chi^{A_{1,1}}_{1/4}\!\left({\tau+24n\over 7}\right)~,
\eea
which are precisely the Harvey-Wu $\mathsf T_7$ relations. 

To summarize, the Harvey-Wu relation in (\ref{eq:HarveyWuexample}) is simply the result of self-duality 
\bea
(E_{7,1} \,\,\mathrm{CFT}) / \langle g \rangle \simeq (E_{7,1} \,\,\mathrm{CFT})
\eea
under the asymmetric orbifold by the order-seven symmetry $g$ described above.  The examples that follow use several variations of this construction. For the Lee-Yang relations, we gauge internal symmetries of $G_{2,1}$ and $F_{4,1}$. Cyclic permutations in three Niemeier lattice theories give various relations for their commutants, as discussed in Section \ref{sec:PermutationCosets}. For Ising, we use permutations of free fermions, a rotation of the $E_{8,2}$ currents, and an order-47 symmetry of the Baby Monster theory. 

\subsection{Further relations}

The self-duality argument also gives relations outside the original Harvey-Wu construction. We collect some resulting identities here. First, the conformal embedding $D_{4,1}\otimes D_{4,1}\subset E_{8,1}$ gives the degree-two relation
\bea
\chi^{E_{8,1}}_0(\tau)&=&\chi^{D_{4,1}}_0(2\tau)-\chi^{D_{4,1}}_v(2\tau)\no\\
&&{}+2\chi^{D_{4,1}}_v\left({\tau\over2}\right)+2\omega\chi^{D_{4,1}}_v\left({\tau+1\over2}\right)~,
\eea
where $\omega=e^{2\pi i/3}$, and $v$ denotes the vector module of $D_{4,1}$. Second, the embedding $A_{2,1}^{\otimes3}\subset E_{6,1}$ gives the degree-three relations,
\bea
\chi^{E_{6,1}}_0(\tau)&=&\chi^{A_{2,1}}_0(3\tau)-\chi^{A_{2,1}}_{\mathbf3}(3\tau)+\sum_{j=0}^2i^j\chi^{A_{2,1}}_{\mathbf3}\left({\tau+j\over3}\right)~,\no\\
\chi^{E_{6,1}}_{\mathbf{27}}(\tau)&=&\chi^{E_{6,1}}_{\overline{\mathbf{27}}}(\tau)=\sum_{j=0}^2(i\omega)^j\chi^{A_{2,1}}_{\mathbf3}\left({\tau+j\over3}\right)~,
\eea
which are direct analogs of (\ref{eq:E7A1finalrelation}).
Here $\mathbf3$ is the fundamental $A_{2,1}$ module. In both cases, a permutation of the identical current-algebra factors is combined with a transformation of the extension sectors. Gauging its action on the left-movers returns the original diagonal CFT, as explained in Sections \ref{sec:DfourEeight} and \ref{sec:AtwoEsix}. The conductors of $D_{4,1}$ and $A_{2,1}$, namely the order of the $T$-matrices, are six and twelve, respectively. Since the degrees two and three are not coprime to these conductors, these identities lie outside the definition of the Harvey-Wu operators.

There are also examples involving minimal-model characters. Take two copies of the Ising model $M(3,4)$, and combine their interchange with a spin flip in one copy. The resulting symmetry has order four, and gauging its action on both chiralities returns the same theory.  This gives
\bea 
\label{eq:IsingPairRelations}
(\chi^{\mathrm{Ising}}_0(\tau))^2-(\chi^{\mathrm{Ising}}_{1/2}(\tau))^2
&=&\chi^{\mathrm{Ising}}_0(2\tau)-\chi^{\mathrm{Ising}}_{1/2}(2\tau)~,\no\\
\chi^{\mathrm{Ising}}_0(\tau)\chi^{\mathrm{Ising}}_{1/16}(\tau)
&=&{1\over2}\left[\chi^{\mathrm{Ising}}_{1/16}\left({\tau\over2}\right)+e^{-\pi i/24}\chi^{\mathrm{Ising}}_{1/16}\left({\tau+1\over2}\right)\right]~,\no\\
\chi^{\mathrm{Ising}}_{1/2}(\tau)\chi^{\mathrm{Ising}}_{1/16}(\tau)
&=&{1\over2}\left[\chi^{\mathrm{Ising}}_{1/16}\left({\tau\over2}\right)-e^{-\pi i/24}\chi^{\mathrm{Ising}}_{1/16}\left({\tau+1\over2}\right)\right]~.
\eea
The subscripts are conformal weights. These Ising character identities were discussed as replication identities in \cite{Bantay:2016nkw}, where they were already related to symmetries of the permutation orbifold of two Ising models. Section \ref{sec:IsingPair} gives an alternative derivation from self-duality under $\mathbb Z_4$ gauging.\footnote{Identities involving covering tori can also arise when gauging changes the theory. For example, the known identity $\chi^{M(5,6)}_0(\tau)-\chi^{M(5,6)}_3(\tau)=\chi^{M(2,5)}_{-1/5}(2\tau)$ \cite{Bytsko:1998tb,Feingold:2013toa} can be understood by gauging charge conjugation in the three-state Potts theory, which takes it to diagonal $M(5,6)$. We will not discuss identities of this type here.}

\paragraph{Organization:} We begin in Section \ref{sec:criterion} by developing the general argument and reviewing the Harvey-Wu construction. Section \ref{sec:HarveyWu} then applies this logic to various one-, two-, and three-character examples. Section \ref{sec:generalized} gives additional relations outside the Harvey-Wu construction.
Appendix \ref{app:OrbifoldCalculations} collects the various lattice constructions, modular calculations, and McKay-Thompson identities used in the main text.

\section{General framework}
\label{sec:criterion}

 The introductory example combined two ingredients: a symmetry-inserted trace expressed in terms of characters on covering tori, and self-duality under gauging. We first explain the covering geometry in an ordinary permutation orbifold. We then formulate the self-duality argument for general cyclic symmetries of prime order and specialize to asymmetric actions to extract character identities. 

\subsection{Permutation orbifolds and covering tori}
\label{sec:PermutationCovers}

We begin by reviewing some standard facts about permutation orbifolds. Consider a bosonic theory $\cA$ with equal left and right central charges $c$, and form two identical copies. Let $g$ exchange the copies. In this subsection the exchange acts on both left- and right-movers, so this is an ordinary symmetric permutation orbifold. We will turn to asymmetric actions in subsequent subsections.

Write the original torus as $z\sim z+1\sim z+\tau$, with the first identification defining the spatial circle. As in the introduction, $Z^s_r$ denotes the partition function with twist $g^r$ and insertion $g^s$. Upon crossing $g$, a field in one copy is taken to the other copy. We can therefore replace the two fields on the original torus by a single field on a surface made by joining the two copies. This is the covering-surface description of a permutation orbifold \cite{Bantay:1997ek}.

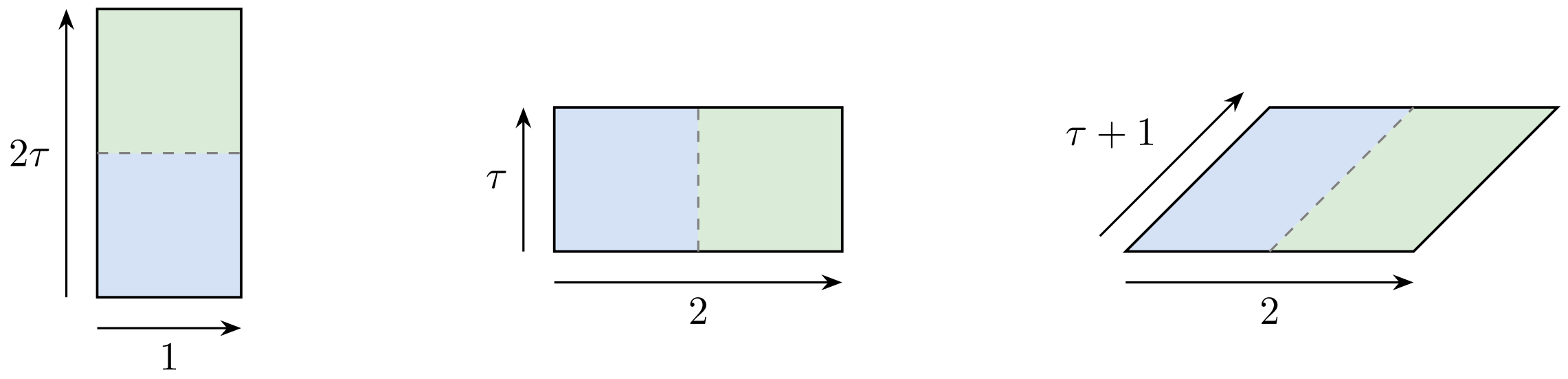
\begin{figure}[t]
\centering
\begin{tikzpicture}[x=0.9cm,y=0.9cm,font=\small,>=Stealth,scale=1.1]
\definecolor{coversblue}{RGB}{213,226,245}
\definecolor{coversgreen}{RGB}{218,235,216}
\tikzset{dedge/.style={line width=0.7pt},dseam/.style={dashed,gray,line width=0.6pt},period/.style={->,line width=0.6pt}}
\begin{scope}[xshift=0.5cm]
\fill[coversblue] (0,0) rectangle (1.4,1.4);
\fill[coversgreen] (0,1.4) rectangle (1.4,2.8);
\draw[dedge] (0,0) rectangle (1.4,2.8);
\draw[dseam] (0,1.4)--(1.4,1.4);
\draw[period] (0,-0.3)--node[below] {$1$}(1.4,-0.3);
\draw[period] (-0.3,0)--node[left] {$2\tau$}(-0.3,2.8);
\end{scope}
\begin{scope}[xshift=4.5cm, yshift=0.4cm]
\fill[coversblue] (0,0) rectangle (1.4,1.4);
\fill[coversgreen] (1.4,0) rectangle (2.8,1.4);
\draw[dedge] (0,0) rectangle (2.8,1.4);
\draw[dseam] (1.4,0)--(1.4,1.4);
\draw[period] (0,-0.3)--node[below] {$2$}(2.8,-0.3);
\draw[period] (-0.3,0)--node[left] {$\tau$}(-0.3,1.4);
\end{scope}
\begin{scope}[xshift=9.5cm, yshift=0.4cm]
\fill[coversblue] (0,0)--(1.4,0)--(2.8,1.4)--(1.4,1.4)--cycle;
\fill[coversgreen] (1.4,0)--(2.8,0)--(4.2,1.4)--(2.8,1.4)--cycle;
\draw[dedge] (0,0)--(2.8,0)--(4.2,1.4)--(1.4,1.4)--cycle;
\draw[dseam] (1.4,0)--(2.8,1.4);
\draw[period] (0,-0.3)--node[below] {$2$}(2.8,-0.3);
\draw[period] (-0.25,0.15)--node[above left] {$\tau+1$}(1.15,1.55);
\end{scope}
\end{tikzpicture}
\caption{The three connected double covers of a torus with periods $1$ and $\tau$, drawn with $\tau$ purely imaginary. Each shaded region is a fundamental region of one copy. The arrows give the two periods of the covering torus; dividing the second by the first gives its modular parameter. }
\label{fig:TorusCovers}
\end{figure}

If neither cycle exchanges the copies, the two fields remain independent, and
\bea \label{eq:PermutationDisconnected}
Z^0_0(\tau,\bar\tau)=Z_{\cA}(\tau,\bar\tau)^2~.
\eea
This is a disconnected cover consisting of two copies of the original torus. 
Next consider $Z^1_0$. The spatial circle has its original length, but a field returns to its original copy only after traversing the Euclidean time cycle twice. The covering torus therefore has periods $1$ and $2\tau$, giving $Z_{\cA}(2\tau,2\bar\tau)$; see Figure \ref{fig:TorusCovers}. We can also see this directly in the trace. If $|i\rangle$ is a basis of states of $\cA$, with weights $(h_i,\bar h_i)$, then $g|i,j\rangle=|j,i\rangle$. A diagonal matrix element of the inserted trace vanishes unless $i=j$. Thus
\bea
Z^1_0(\tau,\bar\tau)&=&\Tr_{\cH_{\cA}\otimes\cH_{\cA}}\!\left(gq^{L_0^{(1)}+L_0^{(2)}-c/12}\bar q^{\bar L_0^{(1)}+\bar L_0^{(2)}-c/12}\right)\no\\
&=&\sum_i q^{2(h_i-c/24)}\bar q^{2(\bar h_i-c/24)}=Z_{\cA}(2\tau,2\bar\tau)~,
\eea
where $q=e^{2\pi i\tau}$ and $\bar q=e^{-2\pi i\bar\tau}$.

For $Z^0_1$, the swap instead joins the two spatial circles. Following a field once around space takes it to the other copy, and a second traversal brings it back. The field now lives on a circle of twice the length, while the Euclidean time period is unchanged. The cover has periods $2$ and $\tau$; dividing the two gives modular parameter $\tau/2$. Consequently,
\bea \label{eq:PermutationSpatialTrace}
Z^0_1(\tau,\bar\tau)=Z_{\cA}\left({\tau\over2},{\bar\tau\over2}\right)~.
\eea
The same statement follows from the twisted spectrum: a state of $\cA$ with weight $h$ gives a left-moving twisted weight $h/2+c/16$. Subtracting the product theory's vacuum energy $2c/24$ leaves $(h-c/24)/2$, precisely the exponent in (\ref{eq:PermutationSpatialTrace}). The right-moving calculation is identical.

Finally, in $Z^1_1$ both cycles exchange the copies. On the doubled spatial circle, inserting $g$ translates by one spatial period. The time identification therefore includes this translation, and the two periods are $2$ and $\tau+1$. Equivalently, the insertion multiplies the state of weights $(h,\bar h)$ by $e^{\pi i(h-\bar h)}$. Either description gives
\bea \label{eq:PermutationMixedTrace}
Z^1_1(\tau,\bar\tau)=Z_{\cA}\left({\tau+1\over2},{\bar\tau+1\over2}\right)~.
\eea
These expressions obey the expected modular transformations: $S$ exchanges the temporal and spatial directions, and $T$ takes (\ref{eq:PermutationSpatialTrace}) to (\ref{eq:PermutationMixedTrace}).

The symmetric product orbifold then sums these four backgrounds with a factor of one-half,
\bea \label{eq:SymmetricSquarePartition}
Z_{\cA^{\otimes2}/S_2}(\tau,\bar\tau)&=&{1\over2}Z_{\cA}(\tau,\bar\tau)^2+{1\over2}\left[Z_{\cA}(2\tau,2\bar\tau)+Z_{\cA}\left({\tau\over2},{\bar\tau\over2}\right)+Z_{\cA}\left({\tau+1\over2},{\bar\tau+1\over2}\right)\right]
\no\\
&=& {1\over2}Z_{\cA}(\tau,\bar\tau)^2 + \mathsf{T_2} Z_{\cA}(\tau,\bar\tau)~,
\eea
where in the second line we have rewritten the expression in square brackets in terms of the weight-zero scalar Hecke transform.\footnote{Note that the standard normalization of  the scalar Hecke operator differs from that of the Harvey-Wu operator $\mathsf{T}_p$ used below, which has no overall factor of $1/p$.} Equation (\ref{eq:SymmetricSquarePartition}) is the two-copy instance of the general organization of symmetric-product partition functions in terms of Hecke operators \cite{Bantay:1997ek,Dijkgraaf:1996xw}.

The counting of covering tori is equally concrete for a more general prime number $p$ of copies. Let $g$ be a $p$-cycle. Every nonzero background $(r,s)$ joins all $p$ copies, since any non-identity power of $g$ acts transitively on them. Multiplying both $r$ and $s$ by a nonzero number modulo $p$ only relabels the copies and gives the same cover. Each cover is therefore counted $p-1$ times among the $p^2-1$ nonzero backgrounds, leaving $p+1$ distinct covers. A symmetry insertion gives the cover with parameter $p\tau$, and modular $S$ and $T$ transformations give the remaining parameters,
\bea \label{eq:CoverParameters}
p\tau~,
\hspace{0.4 in}
{\tau+j\over p}~,
\hspace{0.4 in}
j=0,\ldots,p-1~.
\eea

In the symmetric permutation construction described above, both chiralities are evaluated on the covering tori.
In the asymmetric examples of interest below, the covering tori instead occur for the left-moving characters, while the right-moving characters remain on the original torus. We now explain how self-duality under gauging turns those traces into character identities.

\subsection{Twisted partition functions and self-duality}
\label{sec:TwistedPartitionFunctions}

Let $\cB$ be a bosonic CFT with an anomaly-free symmetry $g$ of prime order $p$. We start from a diagonal CFT. For a holomorphic chiral algebra $V$, this means the full theory with partition function $|\chi^V_0|^2$. As before, we use the convention
\bea
Z^s_r(\tau,\bar\tau)=\Tr_{\cH_{g^r}}\!\left(g^s q^{L_0-c_L/24}\bar q^{\bar L_0-c_R/24}\right)~,
\eea
where $q=e^{2\pi i\tau}$, $\bar q=e^{-2\pi i\bar\tau}$, and $r,s\in\ZZ_p$. 

The anomaly-free assumption allows a choice of symmetry insertions consistent with modular transformations 
\bea \label{eq:defectmodular}
Z^s_r(-1/\tau,-1/\bar\tau)&=&Z^{-r}_s(\tau,\bar\tau)~,
\no\\
Z^s_r(\tau+1,\bar\tau+1)&=&Z^{s+r}_r(\tau,\bar\tau)~.
\eea
This is needed for the orbifold sum to define a consistent CFT. For example, applying $T$ $p$ times to a $g$-twisted torus should return the same boundary conditions. A state of left and right conformal weights $h_L,h_R$ acquires the phase $e^{2\pi ip(h_L-h_R)}$, so this requires the level-matching condition  \cite{Vafa:1986wx,Freed:1987qk,Dong:1997ea}
\bea
p(h_L-h_R)\in\ZZ~.
\eea
 Equivalently, this is the spin-selection rule $h_L-h_R\in{1\over p}\ZZ$ for an anomaly-free $\ZZ_p$ symmetry \cite{Chang:2018iay,Lin:2021udi}. For the asymmetric orbifolds considered below, $g$ acts only on the left-movers. If the chiral algebra is holomorphic, the right-movers therefore remain in the vacuum module and have integral conformal weights. The left-moving twisted sector weights differ from their lowest value $h_g$ by multiples of $1/p$, so level matching reduces to $ph_g\in\ZZ$.

The orbifold projects onto invariant states in every sector. Its partition function is therefore (\ref{eq:introorbifold}). If $Z_{\cB/\ZZ_p}=Z_{\cB}$, this formula gives (\ref{eq:introselfduality}). For later convenience, it is useful to average over the choice of generator of each nonzero background and define
\bea
\cZ_\infty:={1\over p-1}\sum_{a=1}^{p-1}Z^a_0~,\hspace{0.4 in} \cZ_j:={1\over p-1}\sum_{a=1}^{p-1}Z^{aj}_a~,
\eea
where $j=0,\ldots,p-1$. Every nonzero pair $(r,s)$ occurs in exactly one of these averages.
Consequently, self-duality gives
\bea \label{eq:averagedselfduality}
Z_{\cB}=\cZ_\infty+\sum_{j=0}^{p-1}\cZ_j~.
\eea
The modular transformations in (\ref{eq:defectmodular}) imply $\cZ_0=S\cZ_\infty$ and $\cZ_j=T^j\cZ_0$, where $S$ and $T$ act by changing both modular parameters. For the permutation symmetry of Section \ref{sec:PermutationCovers}, these $p+1$ terms are the contributions from the covers in (\ref{eq:CoverParameters}), though we do not assume that the symmetry in question is a permutation symmetry here.

\subsection{Extracting a Hecke-like identity}
\label{sec:mainidentity}

Now consider a diagonal theory, with modules defined with respect to its full chiral algebra, and let $\boldsymbol{\chi}^{\cB}$ be the vector with one character for each irreducible module. Distinct modules may have identical characters; we keep a separate entry for each module. The partition function is
\bea
Z_{\cB}=\boldsymbol{\chi}^{\cB} \cdot \overline{\boldsymbol{\chi}^{\cB}}~.
\eea
Because we are considering a symmetry acting only on left-movers, we may write the (averaged) symmetry-inserted partition function $\cZ_\infty$ as 
\bea 
\cZ_\infty=\mathbf{F_\infty}\cdot \overline{\boldsymbol{\chi}^{\cB}}~,
\eea
where the components of $\mathbf{F_\infty}$ are the averaged symmetry-inserted traces for left-movers. Assume that these take the form
\bea \label{eq:covertracehypothesis}
\mathbf{F_\infty}(\tau)=K\boldsymbol{\chi}^{\cA}(p\tau)~,
\eea
for a constant matrix $K$, where $\boldsymbol{\chi}^{\cA}$ is the character vector for some theory $\cA$. We write $S_{\cA},T_{\cA}$ and $S_{\cB},T_{\cB}$ for the corresponding modular matrices, with the central-charge phase included in $T$.
A modular $S$ transformation then gives
\bea 
\cZ_0(\tau,\bar\tau)
=\left[S_{\cB}^{-1}K S_{\cA}\boldsymbol{\chi}^{\cA}(\tau/p)\right]^t\overline{\boldsymbol{\chi}^{\cB}(\tau)}~.
\eea
Indeed, $\boldsymbol{\chi}^{\cA}(-p/\tau)=S_{\cA}\boldsymbol{\chi}^{\cA}(\tau/p)$, while the transformation of the right-moving characters gives the inverse matrix acting on the coefficient vector. 
Applying $T^j$ then gives
\bea \label{eq:allcoefficients}
\mathbf{F}_j(\tau)=T_{\cB}^{-j}S_{\cB}^{-1}K S_{\cA}\boldsymbol{\chi}^{\cA}\!\left({\tau+j\over p}\right)~.
\eea
If gauging preserves the spectrum paired with each right-moving module, the orbifold sum gives\footnote{More precisely, this follows when the right-moving characters are linearly independent. When there are identical characters, we can only determine sums of coefficients, but these can be separated when the calculated coefficient functions also coincide within each group, as in our examples below. }
\bea
\boldsymbol{\chi}^{\cB}={1\over p}\left[\boldsymbol{\chi}^{\cB}+(p-1)\left(\mathbf F_\infty+\sum_{j=0}^{p-1}\mathbf F_j\right)\right]~.
\eea
Rearranging as in (\ref{eq:averagedselfduality}) then gives the following result,

\begin{proposition}
\label{prop:physicalhecke}
Suppose that $g=g_L\otimes 1_R$ is an anomaly-free order-$p$ symmetry of $\cB$, and that $\cB$ is self-dual under gauging $g$. If the symmetry-inserted traces are given by (\ref{eq:covertracehypothesis}), then we have the Hecke-like relation
\bea \label{eq:generalizedcharacteridentity}
\boldsymbol{\chi}^{\cB}(\tau)=K\boldsymbol{\chi}^{\cA}(p\tau)+\sum_{j=0}^{p-1}T_{\cB}^{-j}S_{\cB}^{-1}K S_{\cA}\,\boldsymbol{\chi}^{\cA}\!\left({\tau+j\over p}\right)~.
\eea
\end{proposition}
\noindent
This simple reasoning will allow us to reproduce Harvey-Wu relations, as well as produce several generalizations thereof. 

\subsection{The Harvey-Wu operator}

Let $N$ be the conductor of $\cA$, namely the order of $T_{\cA}$. For $p$ coprime to $N$, let $\bar p$ satisfy $p\bar p=1\pmod N$, and choose $\sigma_p\in SL(2,\ZZ)$ reducing to $\mathrm{diag}(\bar p,p)$ modulo $N$. With the normalization of \cite{Harvey:2018rdc}, the Harvey-Wu operator is
\bea 
\label{eq:ordinaryhecke}
\mathsf T_p\,\boldsymbol{\chi}^{\cA}(\tau):=\rho_{\cA}(\sigma_p)\boldsymbol{\chi}^{\cA}(p\tau)+\sum_{j=0}^{p-1}\boldsymbol{\chi}^{\cA}\!\left({\tau+Nj\over p}\right)~.
\eea
To recover this expression from Proposition \ref{prop:physicalhecke}, suppose that\footnote{The last two conditions mean that the modular representation of $\cB$ is a Galois conjugate of that of $\cA$ \cite{Bantay:2001ni,Harvey:2018rdc,Harvey:2019qzs}. 
}
\bea 
\label{eq:ordinarycompatibility}
K=\rho_{\cA}(\sigma_p)~, \hspace{0.4 in} S_{\cB}=\rho_{\cA}(\sigma_p) S_{\cA}~,\hspace{0.4 in}T_{\cB}=T_{\cA}^{\bar p}~.
\eea
The coefficient of $\boldsymbol{\chi}^{\cA}((\tau+j)/p)$ in (\ref{eq:generalizedcharacteridentity}) is then $T_{\cA}^{-\bar p j}$. Since multiplication by $N$ permutes the residues modulo $p$, choose $j'$ and an integer $k$ with $Nj'=j+pk$. Modulo $N$, this gives $k=-\bar p j$. It follows that
\bea
\label{eq:reindexing}
T_{\cA}^{-\bar p j}\boldsymbol{\chi}^{\cA}\!\left({\tau+j\over p}\right)=\boldsymbol{\chi}^{\cA}\!\left({\tau+Nj'\over p}\right)~.
\eea
The sum in (\ref{eq:generalizedcharacteridentity}) is therefore the second term of (\ref{eq:ordinaryhecke}), and
\bea
\boldsymbol{\chi}^{\cB}=\mathsf T_p\boldsymbol{\chi}^{\cA}~.
\eea
Thus Harvey-Wu relations fit into the framework mentioned above. However, the Harvey-Wu operators are only defined when $p$ is coprime to $N$, whereas the discussion around (\ref{eq:generalizedcharacteridentity}) does not depend on this. Indeed, we will see examples for which $p$ divides $N$ in Section \ref{sec:generalized}. 

\section{The Harvey-Wu examples}
\label{sec:HarveyWu}

We now apply the orbifold argument to reproduce examples from \cite{Harvey:2018rdc}, organized by the number of independent unflavored characters. Every example starts from a diagonal CFT and gauges a symmetry of the left-movers.

\subsection{Theories with one independent character}

We begin with the degree-two one-character relations of \cite{Harvey:2018rdc},
\bea
\label{eq:HWonechar}
\chi^{E_{8,1}^{\otimes2}}_0=\chi^{D_{16}^{+}}_0=\mathsf T_2\chi^{E_{8,1}}_0~.
\eea
Here $D_{16}^{+}$ denotes the lattice chiral algebra usually called $\mathrm{Spin}(32)/\ZZ_2$. We pair each chiral algebra with an identical right-moving copy and gauge a symmetry of the left-movers. 

Let us first start from the diagonal $E_{8,1}^{\otimes2}$ CFT and gauge the symmetry $g$ exchanging its two left-moving factors. The identity $\Tr_{\cH\otimes\cH}(g\,A\otimes A)=\Tr_{\cH}(A^2)$ encountered above allows us to easily compute the symmetry-inserted trace, and modular transformations give the remaining twisted-twined partition functions,
\bea
Z^0_0(\tau)&=&|\chi^{E_{8,1}}_0(\tau)|^4~,\no\\
Z^1_0(\tau)&=&\chi^{E_{8,1}}_0(2\tau) \,(\overline{\chi^{E_{8,1}}_0(\tau) })^2 ~,\no\\
Z^0_1(\tau)&=&\chi^{E_{8,1}}_0(\tau/2)\,(\overline{\chi^{E_{8,1}}_0(\tau) })^2 ~,\no\\
Z^1_1(\tau)&=&e^{4\pi i/3}\chi^{E_{8,1}}_0((\tau+1)/2)\,(\overline{\chi^{E_{8,1}}_0(\tau) })^2 ~.
\eea
The phase in the last line comes from the $T$ transformations of the right-movers and makes the twisted-sector projection ${1\over2}(Z^0_1+Z^1_1)$ retain integer conformal weights. Indeed, the twisted vacuum has weight $1/2$ and is projected out, while the states of weight one are retained.

The orbifolded theory can be identified from these currents. The untwisted invariant currents generate diagonal $E_{8,2}$. In the twisted sector, an $E_{8,1}$ state of weight $h$ has weight $1/2+h/2$, so its 248 currents give 248 additional weight-one states. Together they reconstruct $E_{8,1}\oplus E_{8,1}$, as shown in the original exchange-orbifold construction of \cite{Forgacs:1988iw}. The theory is therefore self-dual under orbifolding, and hence
\bea
|\chi^{E_{8,1}}_0(\tau)|^4 &=& {1\over 2} \left(\chi^{E_{8,1}}_0(\tau)^2 + \chi^{E_{8,1}}_0(2\tau)+\chi^{E_{8,1}}_0(\tau/2)\right. 
\\
&\vphantom{.}& \hspace{1.5in}\left. +e^{4\pi i/3}\chi^{E_{8,1}}_0((\tau+1)/2) \right) (\overline{\chi^{E_{8,1}}_0(\tau) })^2~,\no
\eea
from which we read off\footnote{Since $\chi^{E_{8,1}}_0(\tau+1)=e^{-2\pi i/3}\chi^{E_{8,1}}_0(\tau)$, the final covering term can also be written as $\chi^{E_{8,1}}_0((\tau+3)/2)$, as in the usual Harvey-Wu operator.}
\bea
\left(\chi^{E_{8,1}}_0(\tau)\right)^2\,\,=\,\,\chi^{E_{8,1}}_0(2\tau)+\chi^{E_{8,1}}_0(\tau/2)+e^{4\pi i/3}\chi^{E_{8,1}}_0((\tau+1)/2) \,\,=\,\, \mathsf T_2\chi^{E_{8,1}}_0(\tau)~.
\eea
The Hecke relation in (\ref{eq:HWonechar}) thus follows straightforwardly from self-duality under gauging the left-moving permutation symmetry of $E_{8,1}^{\otimes2}$.  

We can also obtain this relation by starting from the diagonal $D_{16}^{+}$ theory. In this case we use two order-two symmetries, acting on left-moving momentum vertex operators as
\bea
g_\pm:\,\,e^{i\lambda\cdot X}\longmapsto e^{2\pi i u_\pm\cdot\lambda}e^{i\lambda\cdot X}~, \hspace{0.4 in}u_+={1\over 4}(1^{16})~,\hspace{0.4 in}u_-={1\over 2}(1^8,0^8)~.
\eea
The superscripts in these vectors indicate repeated entries. Since $2u_\pm\in D_{16}^{+}$, the symmetries have order two. Their twisted sector weights are $1/2$ and $1$, respectively. In both cases, twice the weight is integral, so the symmetry can be gauged.

The full partition functions with one symmetry insertion have the form $Z^1_{0,\pm}=F_\pm(\tau)\overline{\chi^{D_{16}^{+}}_0(\tau)}$, where the left-moving coefficients are
\bea
F_\pm(\tau)=\Tr_{\cH_{D_{16}^{+}}}\!\left(g_\pm q^{L_0-2/3}\right)~.
\eea
They can be calculated using the 16 complex fermions describing $D_{16,1}$ and its spinor extension. The first symmetry rotates each fermion by $i$. The second changes the sign of eight fermions, and has vanishing trace on the spinor sector. The resulting expressions are
\bea
F_+(\tau)&=&\chi^{D_{8,1}}_0(2\tau)+\chi^{D_{8,1}}_v(2\tau)+2\chi^{D_{8,1}}_s(2\tau)~,\no\\
F_-(\tau)&=&\chi^{D_{8,1}}_0(2\tau)-\chi^{D_{8,1}}_v(2\tau)~,
\eea
where $0,v,s$ label the vacuum, vector, and spinor modules of $D_{8,1}$. Since $E_{8,1}$ is the spinor extension of $D_{8,1}$, their average is
\bea \label{eq:RankSixteenAverageTrace}
{F_+(\tau)+F_-(\tau)\over2}=\chi^{E_{8,1}}_0(2\tau)~.
\eea

To show self-duality, note that each gauging produces another even unimodular rank-sixteen left-moving lattice theory. Every such lattice has theta function $E_4^2$, so its character is $E_4^2/\eta^{16}=\chi^{D_{16}^{+}}_0$, where $E_4$ is the weight-four Eisenstein series and $\eta$ is the Dedekind eta function. Since the right-movers are unchanged, both orbifold partition functions equal the original one. Thus we have $Z_{D_{16}^{+}} = Z_{D_{16}^{+}/\langle g_\pm\rangle}$, or more explicitly
\bea
\left|\chi^{D_{16}^{+}}_0(\tau)\right|^2={1\over2}\left[\chi^{D_{16}^{+}}_0(\tau)+F_\pm(\tau)+F_\pm(-1/\tau)+e^{4\pi i/3}F_\pm(-1/(\tau+1))\right]\overline{\chi^{D_{16}^{+}}_0(\tau)}~,
\eea
the phase on the right again coming from the $T$ transformation of the right-moving character. Comparing its coefficients, moving the untwisted term to the left-hand side, and averaging the two resulting identities using (\ref{eq:RankSixteenAverageTrace}) gives
\bea
\chi^{D_{16}^{+}}_0(\tau)\,\,=\,\,\chi^{E_{8,1}}_0(2\tau)+\chi^{E_{8,1}}_0(\tau/2)+e^{4\pi i/3}\chi^{E_{8,1}}_0((\tau+1)/2) \,\,=\,\,\mathsf T_2\chi^{E_{8,1}}_0(\tau)~,
\eea
which is again the desired result.

\subsection{Theories with two independent characters}

The simplest two-character relations are as follows,
\bea
\boldsymbol{\chi}^{E_{7,1}}=\mathsf T_7\boldsymbol{\chi}^{A_{1,1}}~,
\hspace{0.4 in}
\boldsymbol{\chi}^{G_{2,1}}=\mathsf T_7\boldsymbol{\chi}^{M(2,5)}~,
\hspace{0.4 in}
\boldsymbol{\chi}^{F_{4,1}}=\mathsf T_{13}\boldsymbol{\chi}^{M(2,5)}~.
\eea
The first was derived as an example in the introduction. The next two will use self-duality under gauging a rotation of the left-moving currents. We use vacuum-first ordering for all these character vectors; in particular, $\boldsymbol{\chi}^{M(2,5)}=(\chi^{M(2,5)}_0,\chi^{M(2,5)}_{-1/5})$. 
After explaining these cases, we will discuss three more Hecke relations following from cyclic permutations of current-algebra factors in three different Niemeier lattice CFTs.

\subsubsection{Lee-Yang and $G_{2,1}$}
\label{sec:exceptional}

We begin with $\boldsymbol{\chi}^{G_{2,1}}=\mathsf T_7\boldsymbol{\chi}^{M(2,5)}$. The Lee-Yang model has central charge $-22/5$ and modules of conformal weights $0$ and $-1/5$. Their characters are
\bea \label{eq:LYproducts}
\chi^{M(2,5)}_0(\tau)&=&q^{11/60}\prod_{n=0}^{\infty}{1\over (1-q^{5n+2})(1-q^{5n+3})}~,\no\\
\chi^{M(2,5)}_{-1/5}(\tau)&=&q^{-1/60}\prod_{n=0}^{\infty}{1\over (1-q^{5n+1})(1-q^{5n+4})}~.
\eea
The $G_{2,1}$ theory has central charge $14/5$ and modules $\mathbf1$ and $\mathbf7$, of conformal weights $0$ and $2/5$. 

Let $H_0$ denote the vector of zero modes of the Cartan currents. Normalize long roots to have squared length two, and define $\rho^\vee$ by $(\alpha_i,\rho^\vee)=1$ for every simple root. On a state of weight $\lambda$, $\rho^\vee\cdot H_0$ has eigenvalue $(\rho^\vee,\lambda)$. Now consider the symmetry
\bea 
\label{eq:G2g}
g=\exp\left(2\pi i\,{\rho^\vee\cdot H_0\over 7}\right)~.
\eea
Under this transformation, the current associated to $\alpha$ acquires the phase $e^{2\pi i(\alpha,\rho^\vee)/7}$. Since $\rho^\vee$ lies in the coroot lattice, $g$ has order seven on both modules.

It is a non-trivial fact that 
\bea \label{eq:exceptionaltwining}
\Tr_{\mathcal H^{G_{2,1}}_{\mathbf1}}\left(gq^{L_0-7/60}\right)&=&\chi^{M(2,5)}_{-1/5}(7\tau)~,\no\\
\Tr_{\mathcal H^{G_{2,1}}_{\mathbf7}}\left(gq^{L_0-7/60}\right)&=&-\chi^{M(2,5)}_0(7\tau)~.
\eea
All six non-zero powers of $g$ are conjugate and hence give the same trace.
The origin of these identities is as follows. The first ingredient is cancellation between states with different symmetry eigenvalues. The fourteen currents have each seventh root of unity as an eigenvalue twice, so
\bea
\Tr_{\mathfrak g_2}g=2\sum_{a=0}^6e^{2\pi ia/7}=0~.
\eea
This cancellation extends to the current oscillators. A negative current mode at level $n$, with symmetry eigenvalue $e^{2\pi ia/7}$, contributes $1+e^{2\pi ia/7}q^n+e^{4\pi ia/7}q^{2n}+\cdots$ when we sum over its occupation number. There are two such modes for each $a=0,\ldots,6$. Before imposing the affine null relations, the oscillator contribution is therefore
\bea
\prod_{n=1}^{\infty}\prod_{a=0}^{6}{1\over(1-e^{2\pi ia/7}q^n)^2}=\prod_{n=1}^{\infty}{1\over(1-q^{7n})^2}~.
\eea
We see that the seven eigenvalues combine into $1-q^{7n}$, explaining the appearance of $7\tau$. The null relations remove further states. The complete traces are fixed by their modular transformations and lowest twisted energies, giving (\ref{eq:exceptionaltwining}); the full calculation is given in Appendix \ref{app:exceptionalproof}.

From the above, we have that 
\bea
Z^1_0=\chi^{M(2,5)}_{-1/5}(7\tau)\overline{\chi^{G_{2,1}}_0(\tau)}-\chi^{M(2,5)}_0(7\tau)\overline{\chi^{G_{2,1}}_{\mathbf7}(\tau)}~,
\eea
and using the modular matrices in our conventions, namely
\bea 
\label{eq:G2F4Smatrix}
S_{M(2,5)}={2\over \sqrt5}\begin{pmatrix}-\sin(2\pi/5)&\sin(\pi/5)\\\sin(\pi/5)&\sin(2\pi/5)\end{pmatrix}~,
\hspace{0.2 in}
S_{G_{2,1}}=S_{F_{4,1}}=\begin{pmatrix}0&1\\-1&0\end{pmatrix}S_{M(2,5)}~,
\eea
gives us the S-transformed result,
\bea
Z^0_1=\chi^{M(2,5)}_0(\tau/7)\overline{\chi^{G_{2,1}}_0(\tau)}+\chi^{M(2,5)}_{-1/5}(\tau/7)\overline{\chi^{G_{2,1}}_{\mathbf7}(\tau)}~.
\eea
The other modular images follow by applying $T^j$ to $Z^0_1$; the right-moving characters contribute the factors $e^{2\pi i7j/60}$ and $e^{-2\pi i17j/60}$.

It remains to show that gauging returns $G_{2,1}$. Similarly to the introductory example, use the conformal embedding \cite{Schellekens:1986mb},
\bea \label{eq:exceptionalembedding}
G_{2,1}\otimes F_{4,1}\subset E_{8,1}~,
\eea
with branching
\bea
\chi^{E_{8,1}}_0=\chi^{G_{2,1}}_0\chi^{F_{4,1}}_0+\chi^{G_{2,1}}_{\mathbf7}\chi^{F_{4,1}}_{\mathbf{26}}~.
\eea
Extend $g$ to $E_{8,1}$ by leaving $F_{4,1}$ fixed. The twisted sector weight is $1/7$, so $7h_g=1$ and level-matching allows for a consistent orbifold. The orbifolded theory is a holomorphic theory of central charge eight, so it must again be $E_{8,1}$. The commutant of the surviving $F_{4,1}$ has the two untwisted Cartan currents and two currents from each of the six twisted sectors, giving fourteen currents in total. Together with its central charge $14/5$, this identifies the commutant as $G_{2,1}$; see Appendix \ref{app:exceptionalproof} for more details.
This establishes self-duality under the asymmetric orbifold.

It is now straightforward to understand the Hecke relation. Since $7^{-1}=43\pmod{60}$ and $T_{G_{2,1}}=T_{M(2,5)}^{43}$, the factors from the right-moving characters can be absorbed by reindexing the covering terms, as in (\ref{eq:reindexing}). Reading off the coefficients of the two right-moving characters then gives
\bea
\chi^{G_{2,1}}_0(\tau)&=&\chi^{M(2,5)}_{-1/5}(7\tau)+\sum_{j=0}^6\chi^{M(2,5)}_0\left({\tau+60j\over 7}\right)~,\no\\
\chi^{G_{2,1}}_{\mathbf7}(\tau)&=&-\chi^{M(2,5)}_0(7\tau)+\sum_{j=0}^6\chi^{M(2,5)}_{-1/5}\left({\tau+60j\over 7}\right)~.
\eea
These are the two components of the desired Hecke relation. We have thus understood this relation as a simple consequence of self-duality of $G_{2,1}$ under gauging the left-moving symmetry (\ref{eq:G2g}). 

\subsubsection{Lee-Yang and $F_{4,1}$}
\label{sec:LeeYangF4}

The second relation is $\boldsymbol{\chi}^{F_{4,1}}=\mathsf T_{13}\boldsymbol{\chi}^{M(2,5)}$. The theory $F_{4,1}$ has central charge $26/5$ and modules $\mathbf1$ and $\mathbf{26}$, with conformal weights $0$ and $3/5$. We repeat the construction above, now using the left-moving rotation
\bea 
g=\exp\left(2\pi i{\rho^\vee\cdot H_0\over 13}\right)~,
\hspace{0.4 in}
{\|\rho^\vee\|^2\over 2}=39~,
\hspace{0.4 in}
(\lambda_{\mathbf{26}},\rho^\vee)=8~.
\eea
The same reasoning as before gives
\bea \label{eq:FfourTwining}
\Tr_{\mathcal H^{F_{4,1}}_{\mathbf1}}\left(gq^{L_0-13/60}\right)&=&\chi^{M(2,5)}_{-1/5}(13\tau)~,\no\\
\Tr_{\mathcal H^{F_{4,1}}_{\mathbf{26}}}\left(gq^{L_0-13/60}\right)&=&-\chi^{M(2,5)}_0(13\tau)~.
\eea
Indeed, each thirteenth root of unity occurs four times among the fifty-two currents, so the current trace vanishes, and the oscillator product before imposing null relations becomes $\prod_{n\geq1}(1-q^{13n})^{-4}$; this is the mechanism for the argument $13\tau$. Null relations then reduce this to the claimed result.
The corresponding partition functions are
\bea
Z^1_0&=&\chi^{M(2,5)}_{-1/5}(13\tau)\overline{\chi^{F_{4,1}}_0(\tau)}-\chi^{M(2,5)}_0(13\tau)\overline{\chi^{F_{4,1}}_{\mathbf{26}}(\tau)}~,\no\\
Z^0_1&=&\chi^{M(2,5)}_0(\tau/13)\overline{\chi^{F_{4,1}}_0(\tau)}+\chi^{M(2,5)}_{-1/5}(\tau/13)\overline{\chi^{F_{4,1}}_{\mathbf{26}}(\tau)}~.
\eea

To establish self-duality under gauging, extend the rotation to $E_{8,1}$ through (\ref{eq:exceptionalembedding}), now leaving $G_{2,1}$ fixed. The holomorphic twisted sector weight is $3/13$. Since $13h_g=3\in \ZZ$, the gauging is consistent, and the holomorphic orbifold again gives $E_{8,1}$. The commutant of $G_{2,1}$ has four untwisted currents and four from each nontrivial twisted sector,
\bea
\dim\left(\mathrm{Com}(G_{2,1},E_{8,1}/\ZZ_{13})\right)_1=4+12\cdot4=52~.
\eea
At central charge $26/5$, this identifies the algebra as $F_{4,1}$, completing the argument for self-duality; details are given in Appendix \ref{app:exceptionalproof}.

Finally, $13^{-1}=37\pmod{60}$ and $T_{F_{4,1}}=T_{M(2,5)}^{37}$. Writing down the self-duality equation, reading off the coefficients of the right-moving characters, and using the reindexing (\ref{eq:reindexing}) gives
\bea
\chi^{F_{4,1}}_0(\tau)&=&\chi^{M(2,5)}_{-1/5}(13\tau)+\sum_{j=0}^{12}\chi^{M(2,5)}_0\left({\tau+60j\over 13}\right)~,\no\\
\chi^{F_{4,1}}_{\mathbf{26}}(\tau)&=&-\chi^{M(2,5)}_0(13\tau)+\sum_{j=0}^{12}\chi^{M(2,5)}_{-1/5}\left({\tau+60j\over 13}\right)~,
\eea
which is the claimed $\mathsf T_{13}$ relation.

\subsubsection{Permutation orbifolds of lattice commutants}
\label{sec:PermutationCosets}

Three further Harvey-Wu relations involve commutants in holomorphic lattice theories \cite{Gaberdiel:2016zke,Harvey:2018rdc},
\bea \label{eq:PermutationCosetGoals}
\boldsymbol\chi^{\mathrm{Coset}_{23}}=\mathsf T_{23}\boldsymbol\chi^{A_{1,1}}~,\hspace{0.4 in}
\boldsymbol\chi^{\mathrm{Coset}_{22}}=\mathsf T_{11}\boldsymbol\chi^{A_{2,1}}~,\hspace{0.4 in}
\boldsymbol\chi^{\mathrm{Coset}_{20}}=\mathsf T_5\boldsymbol\chi^{D_{4,1}}~.
\eea
The three constructions use different Niemeier lattices, specified by the root systems $A_1^{24}$, $A_2^{12}$, and $D_4^6$ \cite{Schellekens:1992db,Conway:1999dov}. The corresponding holomorphic theories are extensions of $A_{1,1}^{\otimes24}$, $A_{2,1}^{\otimes12}$, and $D_{4,1}^{\otimes6}$ by additional momentum sectors. Their commutants $\mathrm{Coset}_{23}$, $\mathrm{Coset}_{22}$, and $\mathrm{Coset}_{20}$ are obtained by removing one level-one factor. Their two independent characters have lowest weights $(0,7/4)$, $(0,5/3)$, and $(0,3/2)$, respectively.

The character vectors in (\ref{eq:PermutationCosetGoals}) have two, three, and four entries, on both sides of the respective identities. In particular, $A_{2,1}$ has modules $\mathbf1,\mathbf3,\overline{\mathbf3}$, but $\chi^{A_{2,1}}_{\mathbf3}=\chi^{A_{2,1}}_{\overline{\mathbf3}}$ without chemical potentials; similarly, the $v,s,c$ modules of $D_{4,1}$ share one character. Their commutants have the same respective multiplicities. Thus all cases have only two independent characters, which is why these examples are included in the present subsection. 

All three constructions have the same form. Write $R=A_1,A_2,D_4$, with rank $r=1,2,4$ and $p=23,11,5$, respectively, so that $r(p+1)=24$. In the corresponding Niemeier theory with roots $R^{p+1}$, choose an order-$p$ symmetry $g$ that cycles $p$ copies of $R_1$ and fixes the remaining copy pointwise. The chosen cycle preserves the additional momentum sectors in the extension and lifts to an order-$p$ symmetry of the full Niemeier theory. Appendix \ref{app:PermutationCosets} gives this extension and the choice of lift. We gauge the resulting action on the left-movers of the diagonal commutant theory.

The symmetry-inserted trace follows from the permutation. An extension sector contributes only if its labels on the $p$ permuted factors are all equal. Each allowed common module label on the $p$ permuted copies is paired with a unique module label on the remaining, unpermuted $R_1$ copy. The permutation trace over the first $p$ copies gives the single-copy character at $p\tau$, while the unpermuted copy contributes its character at $\tau$. Decomposing with respect to this last copy therefore determines the symmetry-inserted traces of the commutant modules. 
For every $s=1,\ldots,p-1$ this gives\footnote{By slight abuse of notation, we use $R$ to also denote the root lattice of the relevant algebra, with $R^*$ its dual. For the simply laced algebras discussed here, $R^*$ is the weight lattice, and its cosets modulo $R$ label the level-one modules. We label the commutant modules by $a\in R^*/R$, with their pairing to these level-one modules specified in Appendix \ref{app:PermutationCosets}.}
\bea \label{eq:PermutationCosetInsertion}
Z^s_0(\tau,\bar\tau)=\sum_{a\in R^*/R}\chi^{R_1}_{-a}(p\tau)\overline{\chi^{\mathrm{Coset}_{pr}}_a(\tau)}~.
\eea
Here $-a$ denotes the conjugate module. It equals $a$ for $A_1$ and $D_4$, while for $A_2$ it interchanges the two non-vacuum modules. Modular $S$ gives
\bea \label{eq:PermutationCosetSpatial}
Z^0_1(\tau,\bar\tau)=\sum_{a\in R^*/R}\chi^{R_1}_a(\tau/p)\overline{\chi^{\mathrm{Coset}_{pr}}_a(\tau)}~,
\eea
and further modular transformations can be used to obtain the remaining twisted-twined partition functions.

To establish self-duality, first perform the same gauging in the auxiliary Niemeier theory. Its twisted sector weight is $h_g=(p-1)/p$, so $p h_g$ is integral and the gauging is consistent. The untwisted invariant currents form $R_p\oplus R_1$. Each of the $p-1$ nontrivial twisted sectors contributes another $\dim R$ currents. Indeed, its vacuum contribution begins with $q^{-1/p}(1+\dim R\,q^{1/p}+\cdots)$, and no non-vacuum branch contributes at weight one. The resulting numbers of currents are
\bea 
\dim W_1=(p+1)\dim R=72,\ 96,\ 168~,
\eea
where $W$ denotes the holomorphic orbifold algebra.

The quantum $\ZZ_p$ symmetry identifies these algebras more precisely. Since $W$ is holomorphic with central charge twenty-four and $\mathrm{dim} W_1>24$, its weight-one Lie algebra $W_1$ is semisimple \cite{dong2004holomorphic}. Its fixed currents are exactly $R_p\oplus R_1$. An inner action would make the full current algebra have rank $2r$, but no semisimple algebra of that rank has the required dimension. The quantum symmetry must therefore permute $p$ simple factors. Their fixed diagonal is one of the two copies of $R$, forcing the full algebra to be $R^{p+1}$, with all factors at level one. Its rank is twenty-four, so the lattice characterization of \cite{Dong:2002fs} and uniqueness of the corresponding Niemeier lattice identify $W$ with the original theory. The isomorphism can be chosen to fix the extra $R_1$ pointwise. Taking its commutant then identifies the gauged diagonal theory with the original one, including its modules and operator products. Appendix \ref{app:PermutationCosets} gives the details of these identifications.

Finally, the conductors of $A_{1,1}$, $A_{2,1}$, and $D_{4,1}$ are $N=24,12,6$. Since $p=-1\pmod N$ in each case, we may choose $\sigma_p=-I$ in (\ref{eq:ordinaryhecke}). Its action is $\rho_{R_1}(-I)=S_{R_1}^2$, which sends each module to its conjugate. Thus the matrix multiplying $\boldsymbol\chi^{R_1}(p\tau)$ in the Harvey-Wu operator has entries $K_{ab}=\delta_{a,-b}$, precisely as in (\ref{eq:PermutationCosetInsertion}). Conjugation fixes the $A_1$ and $D_4$ labels and interchanges the two non-vacuum $A_2$ labels. The modular data of the commutant are complex conjugate to those of $R_1$, so the compatibility conditions (\ref{eq:ordinarycompatibility}) hold. Applying the orbifold sum to (\ref{eq:PermutationCosetInsertion}) and (\ref{eq:PermutationCosetSpatial}) then gives
\bea 
\chi^{\mathrm{Coset}_{pr}}_a(\tau)=\chi^{R_1}_{-a}(p\tau)+\sum_{j=0}^{p-1}\chi^{R_1}_a\left({\tau+Nj\over p}\right)~,
\eea
which are precisely the three relations in (\ref{eq:PermutationCosetGoals}).

\subsection{Theories with three independent characters}

We now consider the Ising character vector, ordered by conformal weights $(0,1/2,1/16)$. The Harvey-Wu relations to be explained are
\bea \label{eq:IsingExamples}
\boldsymbol{\chi}^{\mathrm{Spin}(p)_1}&=&\mathsf T_p\boldsymbol{\chi}^{\mathrm{Ising}}~,\hspace{0.4 in}p=5,7,11,13,17,19,23~,\no\\
\boldsymbol{\chi}^{E_{8,2}}&=&\mathsf T_{31}\boldsymbol{\chi}^{\mathrm{Ising}}~,\no\\
\boldsymbol{\chi}^{\mathrm{BM}}&=&\mathsf T_{47}\boldsymbol{\chi}^{\mathrm{Ising}}~.
\eea
For $E_{8,2}$ we use the module ordering $(\mathbf1,\mathbf{3875},\mathbf{248})$, and for Baby Monster the ordering by conformal weights $(0,3/2,31/16)$. The ordering for $\mathrm{Spin}(p)_1$ depends on $p$, as specified below. As we will now show, the first family of relations comes from self-duality under gauging a permutation of free fermions, while the last two use self-duality under gauging other internal symmetries.

\subsubsection{Ising and $\mathrm{Spin}(p)_1$}

The first relation in (\ref{eq:IsingExamples}) follows from free fermions. For odd $p$, the theory $\mathrm{Spin}(p)_1$ is the bosonic theory associated with $p$ Majorana fermions. Its vacuum, vector, and spinor modules have conformal weights $0$, $1/2$, and $p/16$, and its central charge is $p/2$. For the seven primes listed above, we will show that there is a self-duality under gauging a cyclic permutation of the left-moving fermions.

We begin with the diagonal CFT and let $g$ cyclically permute the $p$ left-moving fermions, while acting trivially on every right-moving field. To specify its action on the spinor module, we choose the order-$p$ lift of this orthogonal transformation to $\mathrm{Spin}(p)$. Define
\bea
\epsilon_p:=(-1)^{(p^2-1)/8}~.
\eea
In (\ref{eq:IsingExamples}), the $\mathrm{Spin}(p)_1$ character vector is ordered as $(0,v,s)$ when $\epsilon_p=1$, and as $(v,0,s)$ when $\epsilon_p=-1$. 
The inserted traces are
\bea \label{eq:SpinCoverTraces}
\begin{pmatrix}
\Tr_{\cH^{\mathrm{Spin}(p)_1}_{\mathbf1}}gq^{L_0-p/48}\\
\Tr_{\cH^{\mathrm{Spin}(p)_1}_{v}}gq^{L_0-p/48}\\
\Tr_{\cH^{\mathrm{Spin}(p)_1}_{s}}gq^{L_0-p/48}
\end{pmatrix}
&=&
\begin{pmatrix}
\chi^{\mathrm{Ising}}_0(p\tau)\\
\chi^{\mathrm{Ising}}_{1/2}(p\tau)\\
\epsilon_p\,\chi^{\mathrm{Ising}}_{1/16}(p\tau)
\end{pmatrix}~.
\eea
These identities follow directly from the fermion oscillators. At each positive frequency $r$, the cyclic permutation has eigenvalues $e^{2\pi i k/p}$ on the $p$ fermion creation operators. Since each mode can be occupied at most once, their contribution to the inserted trace is $\prod_{k=0}^{p-1}(1+e^{2\pi i k/p}q^r)=1+q^{pr}$. This is the contribution of a single fermion oscillator with $q$ replaced by $q^p$. The even and odd fermion-number projections then give the first two lines. In the Ramond sector, the lift has trace $\prod_{r=1}^{(p-1)/2}2\cos(\pi r/p)=1$, but its $p$-th power is $\epsilon_p$. Choosing the order-$p$ lift then gives the third line. 
All non-identity powers of $g$ give the same three traces.

The $S$ matrix is the Ising $S$ matrix in the ordering used in (\ref{eq:SpinCoverTraces}),
\bea
S_{\mathrm{Spin}(p)_1}=S_{\mathrm{Ising}}
={1\over 2}\begin{pmatrix}1&1&\sqrt2\\1&1&-\sqrt2\\\sqrt2&-\sqrt2&0\end{pmatrix}~.
\eea
Conjugating $\mathrm{diag}(1,1,\epsilon_p)$ by this matrix gives the identity when $\epsilon_p=1$, and exchanges the first two entries when $\epsilon_p=-1$. Thus  the twisted partition functions are
\bea \label{eq:SpinSpatialTrace}
Z^0_1=\left\{ \begin{matrix}
 \chi^{\mathrm{Ising}}_0(\tau/p)\overline{\chi^{\mathrm{Spin}(p)_1}_0}+\chi^{\mathrm{Ising}}_{1/2}(\tau/p)\overline{\chi^{\mathrm{Spin}(p)_1}_{v}}+\chi^{\mathrm{Ising}}_{1/16}(\tau/p)\overline{\chi^{\mathrm{Spin}(p)_1}_{s}} && \epsilon_p = 1
 \\
\chi^{\mathrm{Ising}}_{1/2}(\tau/p)\overline{\chi^{\mathrm{Spin}(p)_1}_0}+\chi^{\mathrm{Ising}}_0(\tau/p)\overline{\chi^{\mathrm{Spin}(p)_1}_{v}}+\chi^{\mathrm{Ising}}_{1/16}(\tau/p)\overline{\chi^{\mathrm{Spin}(p)_1}_{s}}&& \,\,\,\,\,\epsilon_p = -1
\end{matrix} \right.\,\,\,\,
\eea
Further modular transformations supply the remaining twisted-twined partition functions. To check that they define a consistent order-$p$ orbifold, consider each of the three products of a left-moving character and a right-moving character in (\ref{eq:SpinSpatialTrace}). A left-moving state counted by $\chi^{\mathrm{Ising}}_h(\tau/p)$ has weight
\bea
h_L={p\over48}+{h-1/48\over p}~,
\eea
while its paired right-moving module has weight $h_R=0$, $1/2$, or $p/16$. Thus we have $p(h_L-h_R)\in\ZZ$ in all three pairs for the listed primes. 
Thus the symmetry can be consistently gauged.

We now argue that the theory is self-dual under this gauging. To see this, work temporarily in the fermionic theory. The permutation preserves fermion parity, and since $p$ is odd, there is no extra phase coupling the order-$p$ gauging to the bosonization---such a phase would have to have order dividing both $p$ and two. We can therefore perform the cyclic gauging first and return to the bosonic theory afterwards. The invariant untwisted sector contains only the single weight-$1/2$ field $p^{-1/2}\sum_{a=1}^p\psi_a$. However, each nontrivial twisted sector contributes one more weight-$1/2$ field. These fields can be found among the lowest Ising states. The relevant states are listed in Table \ref{tabSpinTwistedFermions}; they are read from the coefficient paired with the right-moving vector module in (\ref{eq:SpinSpatialTrace}). For example, for $p=5$ the vacuum state at Ising level two has twisted weight $5/48+(2-1/48)/5=1/2$. The state is retained by the orbifold projection because, paired with the right-moving vector ground state, it has integral spin \cite{Barron:2013tka}.

\begin{table}[t]
\centering
\begin{tabular}{c c c c}
\toprule
$p$ & $\epsilon_p$ & module & level of the state\\
\midrule
$5$ & $-1$ & $0$ & $2$\\
$7$ & $+1$ & $1/2$ & $2$\\
$11$ & $-1$ & $0$ & $3$\\
$13$ & $-1$ & $0$ & $3$\\
$17$ & $+1$ & $1/2$ & $2$\\
$19$ & $-1$ & $0$ & $2$\\
$23$ & $+1$ & $1/2$ & $0$\\
\bottomrule
\end{tabular}
\caption{The Ising state that becomes a weight-$1/2$ fermion in each nontrivial permutation-twisted sector. The level is measured above the highest-weight state of the indicated Ising module.}
\label{tabSpinTwistedFermions}
\end{table}

Thus, together with the invariant untwisted fermion, we have a total of $p$ real fermions. The sectors with twists $g^r$ and $g^{-r}$ provide conjugate fields, corresponding to two real fermions. Unitarity implies that their weight-$1/2$ operator products are those of free fermions. They generate central charge $p/2$, exhausting the available central charge. Returning to the bosonic theory therefore gives $\mathrm{Spin}(p)_1$ again, with its vacuum, vector, and spinor modules. The right-moving modules are unchanged, so this identifies the asymmetric orbifold with the original theory.

Self-duality and the results in (\ref{eq:SpinCoverTraces}) then give the first relation in (\ref{eq:IsingExamples}). The restriction to $p<24$ comes from the weight-$1/2$ fields needed to recover the free-fermion theory. As we said above, originally there are $p$ independent fermions, but the untwisted projection keeps only their symmetric combination, so the other $p-1$ must come from twisted sectors. But for $p>24$, even the lowest Neveu-Schwarz twisted state has weight $(p^2-1)/(48p)>1/2$, and excitations can only increase this weight. There are therefore no twisted weight-$1/2$ fields to replace the missing fermions, and hence there can be no self-duality under this permutation orbifold.

\subsubsection{Ising and $E_{8,2}$}
\label{sec:IsingEeighttwo}

Our next aim is the relation $\boldsymbol{\chi}^{E_{8,2}}=\mathsf T_{31}\boldsymbol{\chi}^{\mathrm{Ising}}$. The central charge of $E_{8,2}$ is $31/2$, and its three modules have highest weights $\mathbf1$, $\mathbf{3875}$, and $\mathbf{248}$, with conformal weights $0$, $3/2$, and $15/16$. 

We act on the left-moving $E_{8,2}$ theory, keeping its right-movers fixed. Let $\rho$ be the $E_8$ Weyl vector, with roots of squared length two, and let $H_0$ denote the vector of Cartan current zero modes. Take
\bea
\label{eq:E82sym}
g=\exp\!\left({2\pi i\over31}\rho\cdot H_0\right)~,
\hspace{0.4 in}\rho^2=620~.
\eea
The symmetry-inserted traces on these modules satisfy
 \bea 
 \label{eq:E82CoverTraces}
\begin{pmatrix}
\Tr_{\cH^{E_{8,2}}_{\mathbf1}}gq^{L_0-31/48}\\
\Tr_{\cH^{E_{8,2}}_{\mathbf{3875}}}gq^{L_0-31/48}\\
\Tr_{\cH^{E_{8,2}}_{\mathbf{248}}}gq^{L_0-31/48}
\end{pmatrix}
&=&\begin{pmatrix}
\chi^{\mathrm{Ising}}_0(31\tau)\\
\chi^{\mathrm{Ising}}_{1/2}(31\tau)\\
\chi^{\mathrm{Ising}}_{1/16}(31\tau)
\end{pmatrix}~.
\eea
The basic reason for this is the same as for $G_{2,1}$ and $F_{4,1}$ in Sections \ref{sec:exceptional} and \ref{sec:LeeYangF4}.  Namely, when acting on the currents, each of the 31 eigenvalues occurs eight times, so the total trace is $8\sum_{r=0}^{30}e^{2\pi i r/31}=0$, and hence these contributions drop out. Before imposing the null relations, the negative current modes at level $n$ contribute $\prod_{r=0}^{30}(1-e^{2\pi ir/31}q^n)^{-8}=(1-q^{31n})^{-8}$ to the vacuum sector trace, which gives rise to the parameter $31\tau$. Null relations then give the results in (\ref{eq:E82CoverTraces}). 

In more detail, we may identify the results with Ising characters by comparing their modular transformations and their lowest twisted energies.
For a module of highest weight $\lambda$ and lowest conformal weight $h$, the lowest twisted sector energy is $h-\rho\cdot\lambda/31+20/31$ \cite{Lam:2016sss}. Thus
\bea
\begin{array}{c|c|c|c}
\text{module}&h&\rho\cdot\lambda&h^{(g)}\\ \hline
\mathbf1&0&0&20/31\\
\mathbf{3875}&3/2&46&41/62\\
\mathbf{248}&15/16&29&321/496
\end{array}~.
\eea
Each twisted ground state is unique, and excited states have conformal weights larger by positive multiples of $1/31$. Subtracting $31/48$ from the last column gives $-1/1488$, $23/1488$, and $1/744$. These are precisely the leading exponents of $\chi^{\mathrm{Ising}}_0(\tau/31)$, $\chi^{\mathrm{Ising}}_{1/2}(\tau/31)$, and $\chi^{\mathrm{Ising}}_{1/16}(\tau/31)$. Only the first exponent is negative, and even the smallest possible excitation makes it positive.
This agreement then extends to the full characters by modularity. Indeed, the two sides of (\ref{eq:E82CoverTraces}) have the same negative-power terms both before and after the modular $S$ transformation. Their modular transformations also agree, as verified in Appendix \ref{app:E82Lattice}. Their difference is therefore constant, and their nontrivial phases under $T$ force that constant to vanish.

We next consider gauging the symmetry in (\ref{eq:E82sym}).  Appendix \ref{app:E82Lattice} constructs this orbifold explicitly and shows $E_{8,2}$ is self-dual. 
Thus (\ref{eq:E82CoverTraces}), together with the fact that all non-identity powers of $g$ are conjugate and give the same traces, leads to 
\bea
\begin{pmatrix}
\chi^{E_{8,2}}_0(\tau)\\
\chi^{E_{8,2}}_{\mathbf{3875}}(\tau)\\
\chi^{E_{8,2}}_{\mathbf{248}}(\tau)
\end{pmatrix}
&=&\begin{pmatrix}
\chi^{\mathrm{Ising}}_0(31\tau)\\
\chi^{\mathrm{Ising}}_{1/2}(31\tau)\\
\chi^{\mathrm{Ising}}_{1/16}(31\tau)
\end{pmatrix}
+\sum_{j=0}^{30}\begin{pmatrix}
\chi^{\mathrm{Ising}}_0((\tau+48j)/31)\\
\chi^{\mathrm{Ising}}_{1/2}((\tau+48j)/31)\\
\chi^{\mathrm{Ising}}_{1/16}((\tau+48j)/31)
\end{pmatrix}~,
\eea
which is the $\mathsf T_{31}$ relation of \cite{Harvey:2018rdc}.

\subsubsection{Ising and the Baby Monster}
\label{sec:IsingBaby}

We finally consider the relation $\boldsymbol{\chi}^{\mathrm{BM}}=\mathsf T_{47}\boldsymbol{\chi}^{\mathrm{Ising}}$. Start with the diagonal Baby Monster theory, of central charge $47/2$. It has three modules, labeled by their lowest conformal weights $0$, $3/2$, and $31/16$ \cite{Hoehn:2007vut,Lin:2019hks}. Choose an order-47 Baby Monster symmetry $g$ acting with order 47 on all three modules, and gauge its action on the left-movers while leaving the right-movers unchanged.
It is a non-trivial fact that
\bea \label{eq:BabyCoverTraces}
\begin{pmatrix}
\Tr_{\cH^{\mathrm{BM}}_0}\!\left(gq^{L_0-47/48}\right)\\
\Tr_{\cH^{\mathrm{BM}}_{3/2}}\!\left(gq^{L_0-47/48}\right)\\
\Tr_{\cH^{\mathrm{BM}}_{31/16}}\!\left(gq^{L_0-47/48}\right)
\end{pmatrix}
&=&\begin{pmatrix}
\chi^{\mathrm{Ising}}_0(47\tau)\\
\chi^{\mathrm{Ising}}_{1/2}(47\tau)\\
\chi^{\mathrm{Ising}}_{1/16}(47\tau)
\end{pmatrix}~.
\eea
The same result holds for every non-identity power of $g$. Appendix \ref{app:BabyScalar} derives these identities from McKay-Thompson series.

To establish self-duality under gauging, it is useful to place the Baby Monster theory inside the Monster theory. The Monster contains a commuting Ising algebra, with branching
\bea \label{eq:BabyCompletion}
V^\natural&=&\bigl(\cH^{\mathrm{Ising}}_0\otimes\cH^{\mathrm{BM}}_0\bigr)
\oplus\bigl(\cH^{\mathrm{Ising}}_{1/2}\otimes\cH^{\mathrm{BM}}_{3/2}\bigr)
\oplus\bigl(\cH^{\mathrm{Ising}}_{1/16}\otimes\cH^{\mathrm{BM}}_{31/16}\bigr)~.
\eea
The subgroup fixing this Ising algebra is $2.\mathrm{BM}$. Since 47 is odd, the chosen Baby Monster element has a unique lift of order 47 to this double cover. In the Monster, it belongs to class $47A$ or $47B$, which have the same McKay-Thompson series. The Ising factor in (\ref{eq:BabyCompletion}) is left unchanged and serves only as an auxiliary factor in this construction.

For these order-47 classes, the Monster orbifold theorem identifies the gauged theory with the Monster theory itself \cite{Paquette:2017xui}. The $g$-twisted ground state has weight $46/47$, so $47h_g=46\in \ZZ$ and the orbifold is consistent. The commuting Ising algebra survives because $g$ fixes it. All Ising conformal vectors in the Monster are conjugate \cite{Chen:2014aia}, so the isomorphism after gauging can be chosen to fix this algebra pointwise. Decomposing with respect to the unchanged Ising algebra then gives back the three Baby Monster modules in (\ref{eq:BabyCompletion}). With the original right-moving modules, this establishes self-duality of the diagonal Baby Monster theory under asymmetric orbifolding.
This self-duality together with (\ref{eq:BabyCoverTraces}) then gives
\bea
\begin{pmatrix}
\chi^{\mathrm{BM}}_0(\tau)\\
\chi^{\mathrm{BM}}_{3/2}(\tau)\\
\chi^{\mathrm{BM}}_{31/16}(\tau)
\end{pmatrix}
&=&\begin{pmatrix}
\chi^{\mathrm{Ising}}_0(47\tau)\\
\chi^{\mathrm{Ising}}_{1/2}(47\tau)\\
\chi^{\mathrm{Ising}}_{1/16}(47\tau)
\end{pmatrix}
+\sum_{j=0}^{46}\begin{pmatrix}
\chi^{\mathrm{Ising}}_0((\tau+48j)/47)\\
\chi^{\mathrm{Ising}}_{1/2}((\tau+48j)/47)\\
\chi^{\mathrm{Ising}}_{1/16}((\tau+48j)/47)
\end{pmatrix}~,
\eea
which is the $\mathsf T_{47}$ relation of \cite{Harvey:2018rdc}.

\section{More general relations}
\label{sec:generalized}

While coprimality of the conductor with the degree of the cover is needed to recover the standard Harvey-Wu operator, self-duality under gauging can give additional identities without this restriction. We first illustrate this with two examples in which a permutation of identical factors in a conformal embedding is combined with a separate symmetry acting on left-movers. We then consider two Ising models with a symmetry that acts on both chiralities and has order four. All three examples have self-duality as the organizing principle.

\subsection{A degree-two relation between $D_{4,1}$ and $E_{8,1}$}
\label{sec:DfourEeight}

We start with the diagonal $E_{8,1}$ theory, with partition function $|\chi^{E_{8,1}}_0|^2$, and use the conformal embedding $D_{4,1}\otimes D_{4,1}\subset E_{8,1}$ to describe its left-moving states. Each $D_{4,1}$ factor has four irreducible modules: the vacuum, the vector, and the two spinor modules, labeled by $0,v,s,c$. The central charge of each factor is four, and the three non-vacuum modules all have lowest conformal weight $1/2$. Their unflavored characters coincide by triality, although the modules are distinct. The $E_{8,1}$ vacuum contains the tensor products with matching labels in the two factors,
\bea \label{eq:DfourModules}
\mathcal H^{E_{8,1}}_{\mathbf1}=\left(\mathcal H^{D_{4,1}}_0\otimes\mathcal H^{D_{4,1}}_0\right)\oplus\left(\mathcal H^{D_{4,1}}_v\otimes\mathcal H^{D_{4,1}}_v\right)\oplus\left(\mathcal H^{D_{4,1}}_s\otimes\mathcal H^{D_{4,1}}_s\right)\oplus\left(\mathcal H^{D_{4,1}}_c\otimes\mathcal H^{D_{4,1}}_c\right)~.
\eea
At the level of characters,
\bea
\chi^{E_{8,1}}_0(\tau)=\left(\chi^{D_{4,1}}_0(\tau)\right)^2+3\left(\chi^{D_{4,1}}_v(\tau)\right)^2~.
\eea
The multiplicity three records the three extension summands.

Let $t$ interchange the two $D_4$ factors in the left-moving $E_{8,1}$ chiral algebra. The four extension sectors are labeled by the classes in $D_4^*/D_4\simeq\ZZ_2^2$, and fusion adds these labels. Assigning signs that multiply under fusion therefore gives another symmetry of the extension. Now let $\varepsilon$ act as $+1$ on the $(0,0)$ and $(v,v)$ sectors and as $-1$ on the $(s,s)$ and $(c,c)$ sectors, and define $g=\varepsilon t$. Note that $t$ preserves each of the four sectors in (\ref{eq:DfourModules}), so it commutes with the symmetry $\varepsilon$. Consequently, $g$ also satisfies $g^2=1$. This defines an order-two automorphism of the extension, and as we will show below it is non-anomalous. 

Using the standard permutation results, we compute the $g$-inserted trace,
\bea
\Tr_{\mathcal H^{E_{8,1}}_{\mathbf1}}\left(gq^{L_0-1/3}\right)&=&\chi^{D_{4,1}}_0(2\tau)+\chi^{D_{4,1}}_v(2\tau)-\chi^{D_{4,1}}_s(2\tau)-\chi^{D_{4,1}}_c(2\tau)\no\\
&=&\chi^{D_{4,1}}_0(2\tau)-\chi^{D_{4,1}}_v(2\tau)~.
\eea
Modular transformations may then be used to compute the remaining twisted-twined partition functions. 
This gives
\bea \label{eq:DfourBlocks}
Z^0_0&=&\left|\chi^{E_{8,1}}_0(\tau)\right|^2~,\no\\
Z^1_0&=&\left(\chi^{D_{4,1}}_0(2\tau)-\chi^{D_{4,1}}_v(2\tau)\right)\overline{\chi^{E_{8,1}}_0(\tau)}~,\no\\
Z^0_1&=&2\chi^{D_{4,1}}_v(\tau/2)\overline{\chi^{E_{8,1}}_0(\tau)}~,\no\\
Z^1_1&=&2\omega\chi^{D_{4,1}}_v\left((\tau+1)/2\right)\overline{\chi^{E_{8,1}}_0(\tau)}~,
\eea
where $\omega=e^{2\pi i/3}$ comes from $T$ transformation of the right-moving $E_8$ vacuum character.

Note that the left-moving coefficient of $\overline{\chi^{E_{8,1}}_0}$ in $Z^0_1$ begins with $16q^{1/6}$, corresponding to weight $1/2$ at central charge eight. It satisfies the order-two level matching condition $2h\in\ZZ$, and hence this symmetry is non-anomalous. In contrast, the symmetry $t$ without including $\varepsilon$ would have twisted character $2\chi^{D_{4,1}}_0(\tau/2)$ and weight $1/4$, and so is anomalous. The inclusion of $\varepsilon$ is thus crucial for obtaining a Hecke-like relation. 

We may now gauge $g$, giving a holomorphic left-moving chiral algebra of central charge eight, which must again be the $E_{8,1}$ VOA. The diagonal CFT is likewise self-dual. Substituting (\ref{eq:DfourBlocks}) into the self-duality equation and removing the overall factor of $\overline{\chi^{E_{8,1}}_0}$ gives the Hecke-like relation
\bea
\chi^{E_{8,1}}_0(\tau)=\chi^{D_{4,1}}_0(2\tau)-\chi^{D_{4,1}}_v(2\tau)+2\chi^{D_{4,1}}_v(\tau/2)+2\omega\chi^{D_{4,1}}_v\left((\tau+1)/2\right)~.
\eea
Note that the conductor of $D_{4,1}$ is six, so the degree two is not coprime to it, and hence this is not an example that could be constructed using Harvey-Wu operators.

\subsection{A degree-three relation between $A_{2,1}$ and $E_{6,1}$}
\label{sec:AtwoEsix}

Next we will exhibit a relationship between the characters of  $E_{6,1}$ and  $A_{2,1}$. The  three $A_{2,1}$ modules have labels $\mathbf1$, $\mathbf3$, and $\overline{\mathbf3}$. Their central charge is two, and the non-vacuum modules have weight $1/3$ and equal unflavored characters. The leading terms are
\bea
\chi^{A_{2,1}}_0(\tau)=q^{-1/12}(1+8 q + \cdots)~,
\hspace{0.4 in}
\chi^{A_{2,1}}_{\mathbf3}(\tau)=\chi^{A_{2,1}}_{\overline{\mathbf3}}(\tau)=q^{1/4}(3+9 q + \cdots)~.
\eea
The conformal extension $A_{2,1}^{\otimes3}\subset E_{6,1}$ has vacuum branching
\bea \label{eq:AtwoModules}
\mathcal H^{E_{6,1}}_{\mathbf1}=\left(\mathcal H^{A_{2,1}}_{\mathbf1}\right)^{\otimes3}\oplus\left(\mathcal H^{A_{2,1}}_{\mathbf3}\right)^{\otimes3}\oplus\left(\mathcal H^{A_{2,1}}_{\overline{\mathbf3}}\right)^{\otimes3}~.
\eea
The $\mathbf{27}$ branches into $(\mathbf1,\mathbf3,\overline{\mathbf3})$, $(\overline{\mathbf3},\mathbf1,\mathbf3)$, and $(\mathbf3,\overline{\mathbf3},\mathbf1)$, and the $\overline{\mathbf{27}}$ into their conjugates. Hence
\bea
\chi^{E_{6,1}}_0(\tau)&=&\left(\chi^{A_{2,1}}_0(\tau)\right)^3+2\left(\chi^{A_{2,1}}_{\mathbf3}(\tau)\right)^3~,\no\\
\chi^{E_{6,1}}_{\mathbf{27}}(\tau)&=&\chi^{E_{6,1}}_{\overline{\mathbf{27}}}(\tau)=3\chi^{A_{2,1}}_0(\tau)\left(\chi^{A_{2,1}}_{\mathbf3}(\tau)\right)^2~,
\eea
with the diagonal partition function given by $|\chi^{E_{6,1}}_0|^2+2|\chi^{E_{6,1}}_{\mathbf{27}}|^2$.

Let $t$ cyclically permute the three $A_2$ factors. Multiply it by a symmetry $\varepsilon$ that acts by $1,\omega,\omega^2$ on the three summands of (\ref{eq:AtwoModules}), and call the result $g$. It is order three. The symmetry-inserted vacuum trace is
\bea \label{eq:AtwoTwining}
\Tr_{\mathcal H^{E_{6,1}}_{\mathbf1}}\left(gq^{L_0-1/4}\right)&=&\chi^{A_{2,1}}_0(3\tau)+\omega\chi^{A_{2,1}}_{\mathbf3}(3\tau)+\omega^2\chi^{A_{2,1}}_{\overline{\mathbf3}}(3\tau)\no\\
&=&\chi^{A_{2,1}}_0(3\tau)-\chi^{A_{2,1}}_{\mathbf3}(3\tau)~.
\eea
The three summands of either non-vacuum module are cyclically permuted, so its inserted trace vanishes. The same results hold with $g$ replaced by $g^2$.

The unflavored $A_{2,1}$ characters transform as
\bea
\chi^{A_{2,1}}_0(-1/\tau)&=&{1\over \sqrt3}\left(\chi^{A_{2,1}}_0(\tau)+2\chi^{A_{2,1}}_{\mathbf3}(\tau)\right)~,\no\\
\chi^{A_{2,1}}_{\mathbf3}(-1/\tau)&=&{1\over \sqrt3}\left(\chi^{A_{2,1}}_0(\tau)-\chi^{A_{2,1}}_{\mathbf3}(\tau)\right)~.
\eea
Thus the $S$ transform of (\ref{eq:AtwoTwining}) is $\sqrt3\chi^{A_{2,1}}_{\mathbf3}(\tau/3)$. Including the transformation of the right-moving characters then gives
\bea \label{eq:AtwoBlocks}
Z^0_0&=&\left|\chi^{E_{6,1}}_0(\tau)\right|^2+2\left|\chi^{E_{6,1}}_{\mathbf{27}}(\tau)\right|^2~,\no\\
Z^s_0&=&\left(\chi^{A_{2,1}}_0(3\tau)-\chi^{A_{2,1}}_{\mathbf3}(3\tau)\right)\overline{\chi^{E_{6,1}}_0(\tau)}~,
\hspace{1.2 in}s=1,2~,\no\\
Z^s_1&=&i^s\chi^{A_{2,1}}_{\mathbf3}\left((\tau+s)/3\right)\left(\overline{\chi^{E_{6,1}}_0(\tau)}+2\omega^s\overline{\chi^{E_{6,1}}_{\mathbf{27}}(\tau)}\right)~,
\hspace{0.4 in}s=0,1,2~,\no\\
Z^s_2&=&Z^{2s}_1~,
\hspace{3.22 in}s=0,1,2~.
\eea
The vacuum and non-vacuum $E_6$ characters have $T$ eigenvalues $-i$ and $e^{5\pi i/6}$, whose conjugates are $i$ and $i\omega$. These are the two phases in the third line. It is periodic under $j\mapsto j+3$ because $\chi^{A_{2,1}}_{\mathbf3}(\tau+1)=i\chi^{A_{2,1}}_{\mathbf3}(\tau)$. The twisted sector weight is $1/3$, paired with right weights zero or $2/3$, so the spins obey level matching and the symmetry can be gauged.

To establish self-duality under the asymmetric orbifold, we use the holomorphic extension $E_{6,1}\times A_{2,1}\subset E_{8,1}$. 
For an explicit lattice realization, let $Q=A_2$ be the root lattice, let $\lambda$ be a fundamental weight, and write $Q_a=Q+a\lambda$ for its three weight-lattice cosets, with $a\in\ZZ_3$. Then
\bea
L_{E_8}=\bigcup_{m,n\in\ZZ_3}\left(Q_m\oplus Q_{m+n}\oplus Q_{m-n}\oplus Q_n\right)~.
\eea
The labels are added modulo three. Every nonzero pair $(m,n)$ gives exactly three non-vacuum factors, of total lowest weight one. The nine cosets form an even lattice of index nine over $Q^{\oplus4}$, and its determinant is $3^4/9^2=1$, identifying it with $E_8$.

The cycle $t(x_1,x_2,x_3,x_4)=(x_3,x_1,x_2,x_4)$ sends the pair $(m,n)$ to $(m-n,n)$. Extend the symmetry $\varepsilon$ to act by $\omega^m$ on this coset. On the three successive images of a coset, the $\varepsilon$ phases multiply to $\omega^{m+(m-n)+(m-2n)}=1$. The combined action consequently has order three, and it fixes the fourth $A_2$ algebra pointwise. On the $E_{6,1}$ vacuum, where $n=0$, it has precisely the grading used in (\ref{eq:AtwoTwining}).

To identify the resulting left-moving algebra, consider the $g$-twisted sector $\mathcal H^{E_{8,1}}_g$ of this auxiliary $E_{8,1}$ extension. Its trace is
\bea
\Tr_{\mathcal H^{E_{8,1}}_g}q^{L_0-1/3}=\chi^{A_{2,1}}_{\mathbf3}(\tau/3)\left(\chi^{A_{2,1}}_0(\tau)+2\chi^{A_{2,1}}_{\mathbf3}(\tau)\right)~.
\eea
The twisted sectors have conformal weight $1/3$, so order-three level matching holds, and the symmetry can be gauged. Gauging produces a holomorphic theory of central charge eight, which must again be $E_{8,1}$. The fourth $A_{2,1}$ factor remains unchanged. Since all level-one $A_2$ embeddings in $E_8$ are related by an $E_8$ symmetry, we can identify this factor with the original one. The algebra commuting with it is then $E_{6,1}$, and the branching of $E_{8,1}$ recovers the three original $E_{6,1}$ modules. Together with the unchanged right movers, this establishes self-duality of the diagonal $E_{6,1}$ theory.

Each of the four non-trivial backgrounds in (\ref{eq:AtwoBlocks}) occurs twice. Self-duality under orbifolding consequently gives
\bea
\chi^{E_{6,1}}_0(\tau)&=&\chi^{A_{2,1}}_0(3\tau)-\chi^{A_{2,1}}_{\mathbf3}(3\tau)+\sum_{j=0}^2i^j\chi^{A_{2,1}}_{\mathbf3}\left({\tau+j\over 3}\right)~,\no\\
\chi^{E_{6,1}}_{\mathbf{27}}(\tau)&=&\sum_{j=0}^2(i\omega)^j\chi^{A_{2,1}}_{\mathbf3}\left({\tau+j\over 3}\right)~.
\eea
The multiplicity two in the diagonal partition function is needed when reading off the second identity. 
Note that the conductor of $A_{2,1}$ is twelve, so the degree three again lies outside of the definition of the Harvey-Wu operators.

\subsection{Two Ising models and a $\ZZ_4$ gauging}
\label{sec:IsingPair}

We now give one final example in which the orbifold is actually symmetric. Take two copies of the diagonal Ising CFT, and let $\eta$ denote spin flip in one copy. It acts as $-1$ on the spin field of weights $(1/16,1/16)$ and as $+1$ on the vacuum and energy sectors. Let $P$ exchange the copies, and define
\bea 
g=(\eta\otimes1)P~,
\hspace{0.4 in}g^2=\eta\otimes\eta~,
\hspace{0.4 in}g^4=1~.
\eea
This symmetry acts on both chiralities. 

The two-copy theory is self-dual under gauging $g$. Indeed, two Ising models lie on the reflection-orbifold branch of the compact boson, which is invariant under the $\ZZ_4$ gauging at every radius.  This self-duality is the familiar $\ZZ_4$ duality at the decoupled-Ising point of the critical Ashkin-Teller model \cite{Yang:1987wk,Diatlyk:2023fwf}. An explicit calculation at the level of the torus partition function can be found in \cite{Thorngren:2021yso}.

The relevant twisted-twined partition functions can be built from the four single-Ising twisted-twined partition functions, 
\bea \label{eq:IsingSpinFlipBlocks}
Z^0_{0,\mathrm{Ising}}&=&|\chi^{\mathrm{Ising}}_0|^2+|\chi^{\mathrm{Ising}}_{1/2}|^2+|\chi^{\mathrm{Ising}}_{1/16}|^2~,\no\\
Z^1_{0,\mathrm{Ising}}&=&|\chi^{\mathrm{Ising}}_0|^2+|\chi^{\mathrm{Ising}}_{1/2}|^2-|\chi^{\mathrm{Ising}}_{1/16}|^2~,\no\\
Z^0_{1,\mathrm{Ising}}&=&|\chi^{\mathrm{Ising}}_{1/16}|^2+\chi^{\mathrm{Ising}}_0\overline{\chi^{\mathrm{Ising}}_{1/2}}+\chi^{\mathrm{Ising}}_{1/2}\overline{\chi^{\mathrm{Ising}}_0}~,\no\\
Z^1_{1,\mathrm{Ising}}&=&|\chi^{\mathrm{Ising}}_{1/16}|^2-\chi^{\mathrm{Ising}}_0\overline{\chi^{\mathrm{Ising}}_{1/2}}-\chi^{\mathrm{Ising}}_{1/2}\overline{\chi^{\mathrm{Ising}}_0}~,
\eea
where the labels represent insertions of the spin-flip symmetry $\eta$.
Now use $Z^s_r$ to denote the two-copy theory twisted-twined partition functions, where $r,s=0,1,2,3$ label insertions of $g$. When both labels are even, neither cycle exchanges the copies, and the amplitude is a product of two single-Ising amplitudes. Otherwise the two copies join into one of the three connected double covers in Figure \ref{fig:TorusCovers}. Following the field twice through the exchange also applies $g^2=\eta\otimes\eta$, so the covering theory carries a spin-flip background. For example,
\bea
Z^1_0(\tau,\bar\tau)=Z^1_{0,\mathrm{Ising}}(2\tau,2\bar\tau)~,
\hspace{0.4 in}
Z^0_1(\tau,\bar\tau)=Z^0_{1,\mathrm{Ising}}(\tau/2,\bar\tau/2)~.
\eea
All sixteen backgrounds are listed in Table \ref{tab:IsingPairCovers}. Each of its last six rows occurs twice. Thus the order of the symmetry is four, but the connected covers have degree two: the subgroup generated by $g^2$ acts internally and does not exchange the sheets.

\begin{table}[t]
\hypersetup{citecolor=red,linkcolor=red,urlcolor=red}
\centering
\renewcommand{\arraystretch}{1.25}
\begin{tabular}{cc}
\toprule
two-copy background & single-Ising expression \\
\midrule
$(2u,2v)$, $u,v=0,1$ & $[Z^v_{u,\mathrm{Ising}}(\tau,\bar\tau)]^2$ \\
$(0,1),(0,3)$ & $Z^1_{0,\mathrm{Ising}}(2\tau,2\bar\tau)$ \\
$(2,1),(2,3)$ & $Z^1_{1,\mathrm{Ising}}(2\tau,2\bar\tau)$ \\
$(1,0),(3,0)$ & $Z^0_{1,\mathrm{Ising}}(\tau/2,\bar\tau/2)$ \\
$(1,2),(3,2)$ & $Z^1_{1,\mathrm{Ising}}(\tau/2,\bar\tau/2)$ \\
$(1,1),(3,3)$ & $Z^0_{1,\mathrm{Ising}}((\tau+1)/2,(\bar\tau+1)/2)$ \\
$(1,3),(3,1)$ & $Z^1_{1,\mathrm{Ising}}((\tau+1)/2,(\bar\tau+1)/2)$ \\
\bottomrule
\end{tabular}
\caption{The sixteen backgrounds for the $\ZZ_4$ symmetry of two Ising models. The first row contains four disconnected covers; the other rows contain the twelve connected covers, with their induced spin-flip boundary conditions. }
\label{tab:IsingPairCovers}
\end{table}

Using self-duality, we obtain the relation
\bea \label{eq:IsingPairOrbifold}
[Z^0_{0,\mathrm{Ising}}(\tau,\bar\tau)]^2
&=&{1\over4}\sum_{u,v=0}^1[Z^v_{u,\mathrm{Ising}}(\tau,\bar\tau)]^2 +{1\over2}(Z^1_{0,\mathrm{Ising}}+Z^1_{1,\mathrm{Ising}})(2\tau,2\bar\tau)\no\\
&&{}+{1\over2}\sum_{j=0}^1(Z^0_{1,\mathrm{Ising}}+Z^1_{1,\mathrm{Ising}})\left({\tau+j\over2},{\bar\tau+j\over2}\right)~.
\eea
Substituting (\ref{eq:IsingSpinFlipBlocks}) into (\ref{eq:IsingPairOrbifold}) and moving the disconnected terms to the left gives
\bea \label{eq:IsingPairNormIdentity}
&&\left|(\chi^{\mathrm{Ising}}_0(\tau))^2-(\chi^{\mathrm{Ising}}_{1/2}(\tau))^2\right|^2
+\left|2\chi^{\mathrm{Ising}}_0(\tau)\chi^{\mathrm{Ising}}_{1/16}(\tau)\right|^2+\left|2\chi^{\mathrm{Ising}}_{1/2}(\tau)\chi^{\mathrm{Ising}}_{1/16}(\tau)\right|^2\no\\
&&\hspace{0.4 in}=\left|\chi^{\mathrm{Ising}}_0(2\tau)-\chi^{\mathrm{Ising}}_{1/2}(2\tau)\right|^2
+2\left|\chi^{\mathrm{Ising}}_{1/16}(\tau/2)\right|^2+2\left|\chi^{\mathrm{Ising}}_{1/16}((\tau+1)/2)\right|^2~.
\eea

Since this gauging acts on both chiralities, we cannot hold the original right-moving modules fixed as in Proposition \ref{prop:physicalhecke}. Instead, the conformal weights separate the three terms in (\ref{eq:IsingPairNormIdentity}). To see this, rewrite its last two terms as the sum of the absolute squares of
\bea 
\chi^{\mathrm{Ising}}_{1/16}(\tau/2)
\mathbin{\pm}e^{-\pi i/24}\chi^{\mathrm{Ising}}_{1/16}((\tau+1)/2)~.
\eea
The phase cancels the phase of the leading term under the shift from $\tau/2$ to $(\tau+1)/2$. The plus sign retains its even descendant levels, and the minus sign retains its odd levels. The three holomorphic expressions on each side of the resulting identity have leading terms
\bea
q^{-1/24}~,
\hspace{0.4 in}2q^{1/48}~,
\hspace{0.4 in}2q^{25/48}~,
\eea
respectively, followed in each case by integer powers of $q$. These three exponents are distinct modulo one. Comparing the coefficient of the lowest antiholomorphic power in each class therefore identifies the corresponding holomorphic expressions. This gives the three relations in (\ref{eq:IsingPairRelations}).

\section*{Acknowledgements}

We thank Xiaoyi Shi, Soichiro Shimamori, and Zhengdi Sun for many useful discussions. The author is supported by JSPS KAKENHI Grant No. 26K17152 and through the Inamori Frontier Program at Kyushu University. As with all research in this era, use was made of AI. Broadly organizing the results of human-AI collaboration into two types,
\begin{description}
\item[  Type I:]  human generated, AI verified
\item[  Type II:]  AI generated, human verified 
\end{description}
the results in the current paper were mostly of Type I. Namely, the basic idea and most examples were first worked out by the author, after which a draft was given to OpenAI's GPT 6 for comments and verification.

\appendix

\section{Details of orbifold calculations}
\label{app:OrbifoldCalculations}

In this appendix we derive the various symmetry-inserted traces and self-dualities used in the main text.

\subsection{Details of the order-seven $E_{7,1}$ orbifold}
\label{app:E7}

This appendix supplies details for the example in the introduction. We first realize the seven-cycle inside $E_{8,1}$, with an extra $A_{1,1}$ factor held fixed. We then check consistency of the orbifold and perform the gauging. The right-movers are unchanged throughout.

\paragraph{The symmetry and its gauging}
Let $\alpha_1,\ldots,\alpha_8$ be orthogonal roots with $(\alpha_i,\alpha_j)=2\delta_{ij}$, and put $R_7=\bigoplus_{i=1}^7\ZZ\alpha_i$. The extension in (\ref{eq:EsevenModuleBranching}) has lattice
\bea \label{eq:EsevenLatticeCosets}
L_7=R_7\cup\bigcup_{j=0}^6\left(R_7+{1\over2}\sum_{i\in I_j}\alpha_i\right)~,
\hspace{0.4 in}I_j=\{1+j,2+j,3+j,5+j\}~,
\eea
where indices are reduced modulo seven to $1,\ldots,7$. Each subset has four elements, any two distinct subsets intersect in two elements, and their symmetric difference is another of the seven subsets. Thus (\ref{eq:EsevenLatticeCosets}) is an even lattice of determinant $2^7/8^2=2$. Its roots are the fourteen vectors $\pm\alpha_i$ and the sixteen vectors ${1\over2}\sum_{i\in I_j}\pm\alpha_i$ for each $j$, giving the 126 roots of $E_7$. Adding $\mu=(\alpha_1+\cdots+\alpha_7)/2$ interchanges the two $A_{1,1}$ modules in each factor and gives the non-vacuum branching in (\ref{eq:EsevenBranching}).

Now adjoin the eighth root and $\lambda=(\alpha_1+\cdots+\alpha_8)/2$. The completion is
\bea \label{eq:EsevenLatticeExtension}
L_8=(L_7\oplus\ZZ\alpha_8)\cup\left(\lambda+L_7\oplus\ZZ\alpha_8\right)~.
\eea
This lattice is even and has determinant $(2\cdot2)/2^2=1$, so it is the $E_8$ lattice. The seven-cycle preserves both terms in (\ref{eq:EsevenLatticeExtension}) and fixes $\alpha_8$. Choose the lift that acts as $+1$ on momentum states whose lattice vectors are fixed by the cycle. This is the standard lift; because the cycle has odd order, its lift also has order seven \cite{vanEkeren:2017scl}. It fixes the eighth $A_{1,1}$ pointwise. It also fixes the state of momentum $\lambda$, determining the plus sign in the non-vacuum trace in (\ref{eq:EsevenTwining}).

The trace of the $g^r$-twisted sector follows by applying $S$ to the trace with $g^r$ inserted. From (\ref{eq:EsevenTwining}) and the completion (\ref{eq:EsevenSpectatorBranching}), the latter is
\bea \label{eq:EsevenParentInsertion}
\Tr_{V_{E_8}}\!\left(g^r q^{L_0-1/3}\right)
&=&\chi^{A_{1,1}}_0(7\tau)\chi^{A_{1,1}}_0(\tau)+\chi^{A_{1,1}}_{1/4}(7\tau)\chi^{A_{1,1}}_{1/4}(\tau)~,
\hspace{0.2 in}r=1,\ldots,6~.
\eea
Applying a modular $S$ transformation to (\ref{eq:EsevenParentInsertion}) acts with the matrix (\ref{eq:A11Smatrix}) on both factors. Since the matrix is orthogonal, the cross terms cancel. The result is then
\bea \label{eq:EsevenParentTwisted}
\Tr_{V_{E_8}(g^r)}q^{L_0-1/3}=\chi^{A_{1,1}}_0(\tau/7)\chi^{A_{1,1}}_0(\tau)+\chi^{A_{1,1}}_{1/4}(\tau/7)\chi^{A_{1,1}}_{1/4}(\tau)~,
\eea
where $V_{E_8}(g^r)$ denotes the twisted state space. In (\ref{eq:EsevenParentTwisted}), the first term begins with $q^{-1/168}q^{-1/24}=q^{-1/21}$, while the second starts at a higher power. The exponent in this trace is $L_0-1/3$, so
\bea
h_{g^r}-{1\over3}=-{1\over21}~,
\hspace{0.4 in}h_{g^r}={2\over7}~.
\eea
Thus all six twisted sectors have positive weight and satisfy $7h_{g^r}=2\in\ZZ$. This establishes consistency under gauging. 

\paragraph{Self-duality}
After gauging, the resulting holomorphic theory has central charge eight, so it is again $E_{8,1}$. The extra $A_{1,1}$ survives at level one. To identify it inside the new $E_8$, conjugate its Cartan generator into a Cartan subalgebra of $E_8$. Its integral charges place it in the $E_8$ coweight lattice, which equals the root lattice, and its level-one normalization gives squared norm two. It is therefore an $E_8$ root. Since all roots are conjugate, the isomorphism to $E_{8,1}$ can be chosen to fix this $A_{1,1}$. Its commutant and its other branching space are then the original pair $(V_{E_7},M_{E_7})$ in (\ref{eq:EsevenSpectatorBranching}).

It remains to check that this holomorphic construction uses the same projection as the asymmetric $E_{7,1}$ orbifold. In the untwisted sector, both constructions project onto invariants under the same action of $g$. In each nontrivial twisted sector, the gauging keeps integral total conformal weight. Since $g$ fixes the spectator currents, their modules are the ordinary $A_{1,1}$ modules of weights $0$ and $1/4$ modulo one. The corresponding $E_7$ weights must therefore lie in $\ZZ$ and $3/4+\ZZ$, respectively. These are exactly the conditions imposed by the right-moving $E_7$ modules in (\ref{eq:EsevenAllTraces}). The spectator-preserving isomorphism therefore identifies the two projected left-moving modules and their products with those of $E_{7,1}$. Pairing with the unchanged right-movers recovers the original diagonal $E_{7,1}$ CFT.

\subsection{Details of the $G_{2,1}$ and $F_{4,1}$ orbifolds}
\label{app:exceptionalproof}

We next give details on the arguments in Sections \ref{sec:exceptional} and \ref{sec:LeeYangF4}. 

\paragraph{Symmetry-inserted traces}

For both $G_{2,1}$ ($p=7$) and $F_{4,1}$ ($p=13$), the transformation $g=\exp(2\pi i\rho^\vee\cdot H_0/p)$ has order $p$ on both modules because $\rho^\vee$ belongs to the coroot lattice. We first show that its nontrivial powers are all conjugate. The coefficients $a_i$ of the highest root in the simple-root basis are $(3,2)$ for $G_2$ and $(2,3,4,2)$ for $F_4$. Their sum is $p-2$, so every root $\alpha$ satisfies $0<|(\alpha,\rho^\vee)|\leq p-2$. Under $g^r$, the corresponding current acquires the phase $e^{2\pi i r(\alpha,\rho^\vee)/p}$. Since $p$ is prime, this phase is nontrivial for $r=1,\ldots,p-1$, and only the Cartan currents remain invariant. After conjugation, any order-$p$ rotation with this property can be described by phases $e^{2\pi i s_i/p}$ on the simple-root currents, where the $s_i$ are positive integers satisfying $\sum_i a_i s_i<p$. Since $\sum_i a_i=p-2$ and every $a_i\geq2$, increasing any $s_i$ above one would violate this inequality. The only possibility is therefore $s_i=1$ for every $i$, which gives $g$ itself. Thus every nontrivial power of $g$ is conjugate to $g$.

Now let $\boldsymbol{\chi}^{\mathcal B}_g$ be the vector of chiral traces $\Tr_{\mathcal H^{\mathcal B}_a}(gq^{L_0-c_{\mathcal B}/24})$ for $\mathcal B=G_{2,1}$ or $F_{4,1}$. Pairing its components with the corresponding right-moving characters gives the full partition function $Z^1_0$. We now show that this vector is equal to $(\chi^{M(2,5)}_{-1/5}(p\tau),-\chi^{M(2,5)}_0(p\tau))$. This is done in two steps: we first show that their modular transformations agree, and then show that their negative powers of $q$ agree at both cusps of $\Gamma_0(p)$. 

For the first step, insert a Cartan chemical potential in the characters, apply a modular transformation, and then set the chemical potential to $\rho^\vee/p$. For $\gamma=\left(\begin{smallmatrix}a&b\\pm&d\end{smallmatrix}\right)\in\Gamma_0(p)$, this gives
\bea
\boldsymbol{\chi}^{\mathcal B}_g(\gamma\tau)=\exp\!\left(-2\pi i md{\|\rho^\vee\|^2\over2p}\right)\rho_{\mathcal B}(\gamma)\boldsymbol{\chi}^{\mathcal B}_{g^d}(\tau)=\rho_{\mathcal B}(\gamma)\boldsymbol{\chi}^{\mathcal B}_g(\tau)~.
\eea
The phase is one because $\|\rho^\vee\|^2/2p$ is one for $G_2$ and three for $F_4$, and conjugacy removes the dependence on $d$. The Lee-Yang expression has the same transformation law,
\bea
K\rho_{M(2,5)}\!\left(\begin{smallmatrix}a&pb\\m&d\end{smallmatrix}\right)K^{-1}=\rho_{\mathcal B}(\gamma)~,
\hspace{0.4 in}K=\begin{pmatrix}0&1\\-1&0\end{pmatrix}~.
\eea
To verify this equation, we use the congruence property of the Lee-Yang modular representation---its action depends on the entries of the modular transformation only modulo the conductor, which in this case is sixty  \cite{Bantay:2001ni}. Since $p$ is invertible modulo sixty, the change of argument and conjugation by $K$ give a representation with matrices $KS_{M(2,5)}$ and $T_{M(2,5)}^{\bar p}$ for the modular generators, where $\bar p=43,37$. These are exactly the $G_{2,1}$ and $F_{4,1}$ matrices given in (\ref{eq:G2F4Smatrix}).

At the cusp $\infty$, both vectors have just one negative power $q^{-p/60}$ in the vacuum component, with coefficient one. To check the cusp at zero, apply $S$ and inspect the twisted sectors. The rotation changes their Hamiltonian to 
\bea
L_0^{(g)}=L_0-{\rho^\vee\cdot H_0\over p}+{\|\rho^\vee\|^2\over2p^2}~.
\eea
Thus a highest-weight state of weight $\lambda$ has energy $h_\lambda-(\lambda,\rho^\vee)/p+\|\rho^\vee\|^2/(2p^2)$. The only root data and resulting weights needed here are,
\begin{center}
\begin{tabular}{c c c c c c}
\toprule
$\mathcal B$ & $p$ & $\|\rho^\vee\|^2/2$ & $(\lambda,\rho^\vee)$ & $h^{(g)}_0$ & $h^{(g)}_\lambda$\\
\midrule
$G_{2,1}$ & $7$ & $7$ & $3$ & $1/7$ & $4/35$\\
$F_{4,1}$ & $13$ & $39$ & $8$ & $3/13$ & $14/65$\\
\bottomrule
\end{tabular}
\end{center}
Here $\lambda$ is the highest weight of $\mathbf7$ or $\mathbf{26}$. Each ground state is unique. 
Subtracting $c/24$ from the last two columns gives $11/60p$ and $-1/60p$. Hence the twisted, non-vacuum trace has the sole negative power $q^{-1/60p}$, again with coefficient one. On the Lee-Yang side, (\ref{eq:G2F4Smatrix}) gives
\bea
\begin{pmatrix}\chi^{M(2,5)}_{-1/5}(-p/\tau)\\-\chi^{M(2,5)}_0(-p/\tau)\end{pmatrix}=S_{\mathcal B}\begin{pmatrix}\chi^{M(2,5)}_0(\tau/p)\\\chi^{M(2,5)}_{-1/5}(\tau/p)\end{pmatrix}~,
\eea
with exactly these leading powers. The two sides therefore agree in every negative power at both cusps. Their difference has weight zero and a finite-image multiplier, so each component is invariant under a finite-index subgroup of the modular group. Since there are no poles, either in the upper half-plane or at the cusps, the maximum principle forces each component to be constant \cite{DHoker:2024xor}. Neither $T$ eigenvalue is one, forcing the constant to vanish. This proves (\ref{eq:exceptionaltwining}) and (\ref{eq:FfourTwining}).

\paragraph{Self-duality}

To identify the result after gauging, extend the rotation to $E_{8,1}$ using $G_{2,1}\otimes F_{4,1}\subset E_{8,1}$, leaving the other factor fixed. The vacuum branch has twisted sector weight $1/7$ or $3/13$, as above. Adding the unchanged factor's weight in the other branch gives $5/7$ or $8/13$, so the vacuum branch is the ground state of the full $E_8$ twisted sector. All nontrivial powers have the same spectrum. The weights are positive and satisfy $ph_g=1$ or $3$, so the orbifold is consistent and must give $E_{8,1}$.

We then identify the commutant of the unchanged factor. Its untwisted invariant currents are the two or four Cartan currents of the rotated algebra. In each nontrivial twisted sector, the holomorphic projection keeps states of integral total conformal weight. The commutant vacuum trace is $\chi^{M(2,5)}_0(\tau/p)$. 
The coefficient of $q^n$ in $q^{-11/60}\chi^{M(2,5)}_0(\tau)$ counts twisted-sector states of conformal weight $h_g+n/p$, since evaluating the character at $\tau/p$ divides the excitation energies by $p$. A current therefore requires $n=6$ for $G_2$ or $n=10$ for $F_4$. The product in (\ref{eq:LYproducts}) gives coefficients two and four at these degrees. The commutants consequently have
\bea
\dim\mathrm{Com}(F_{4,1},E_{8,1}/\ZZ_7)_1&=&2+6\cdot2=14~,\no\\
\dim\mathrm{Com}(G_{2,1},E_{8,1}/\ZZ_{13})_1&=&4+12\cdot4=52~.
\eea
If the current algebra has rank $r$, its $r$ commuting Cartan generators generate $U(1)^r$, with central charge $r$. Unitarity requires the remaining contribution to the central charge to be nonnegative, so $r\leq c$. For the first commutant, $c=14/5$ therefore gives $r\leq2$. The simple Lie algebras of rank at most two are $A_1$, $A_2$, $B_2=C_2$, and $G_2$, with dimensions $3$, $8$, $10$, and $14$. A reducible algebra has at most the six currents of $A_1\oplus A_1$; abelian factors only decrease this maximum. Hence fourteen currents require $G_2$.

For the second commutant, $c=26/5$ gives $r\leq5$. The simple rank-five algebras $A_5$, $D_5$, $B_5$, and $C_5$ have dimensions $35$, $45$, $55$, and $55$. At smaller rank, $F_4$ has dimension $52$, while every other simple algebra has dimension at most $36$. Reducible algebras cannot give $52$: without an $F_4$ factor, the largest possibility is $B_4\oplus A_1$ or $C_4\oplus A_1$, of dimension $36+3=39$. Splitting the rank as $3+2$ gives at most $21+14=35$, and further splitting gives fewer currents. Adding any factor to $F_4$ instead exceeds $52$. Thus fifty-two currents require $F_4$.

The Sugawara central charges $14k/(k+4)$ and $52k/(k+9)$ then force $k=1$, since the available central charges are $14/5$ and $26/5$. These currents exhaust the stress tensor. The other level-one module has nonintegral conformal weight, so it cannot further extend the integer-graded vacuum algebra. The two commutants are therefore precisely $G_{2,1}$ and $F_{4,1}$.

\subsection{Permutation orbifolds of Niemeier commutants}
\label{app:PermutationCosets}

Here we check the permutation-inserted traces and the identification of the unchanged current-algebra factor used in Section \ref{sec:PermutationCosets}. Write $(R,p,r)=(A_1,23,1)$, $(A_2,11,2)$, or $(D_4,5,4)$, and let $L$ be the Niemeier lattice with roots $R^{p+1}$. Its lattice symmetries include an order-$p$ permutation of $p$ components fixing the remaining component \cite{Conway:1999dov}. The fixed component has no diagram symmetry of order $p$, so it is fixed pointwise. Choose the lift that acts as $+1$ on momentum states fixed by the lattice permutation. Since $p$ is odd, this lift and all its non-identity powers have order $p$ \cite{vanEkeren:2017scl}. 

The commutant is the lattice theory of $K=R^\perp\cap L$. As for $R^*$, the dual lattice $K^*$ consists of vectors in the real span of $K$ having integer inner product with every vector of $K$. Its cosets $K^*/K$ label the commutant modules: momenta in the same coset differ by a vector of $K$. There are no additional lattice vectors supported on the fixed component, since the non-vacuum $R_1$ modules have non-integral conformal weight. Unimodularity therefore pairs these module labels with $R^*/R$.
 Because $g$ fixes $R$, it fixes these labels. Acting with $g$ on the left-movers and trivially on the right-movers consequently preserves the diagonal commutant theory.

\paragraph{Symmetry-inserted traces}
Write $a\in R^*/R$ for the module labels of $R_1$. Only sectors with the same label $a$ on all $p$ permuted copies contribute to the trace. For each $a$, there is exactly one such sector in the Niemeier theory; denote its label on the remaining copy by $\phi(a)$. This gives two, three, or four contributing sectors for $A_1$, $A_2$, or $D_4$, respectively. The map $\phi$ is a one-to-one relabeling compatible with fusion and conformal weights modulo one.

Only these sectors contribute to the symmetry-inserted trace. The cyclic trace on the first $p$ factors replaces $\tau$ by $p\tau$, and the chosen lift introduces no additional factors. Hence
\bea
\Tr_{V_L}\!\left(gq^{L_0-1}\right)&=&\sum_{a\in R^*/R}\chi^{R_1}_a(p\tau)\chi^{R_1}_{\phi(a)}(\tau)~,\no\\
\Tr_{V_L(g)}q^{L_0-1}&=&\sum_{a\in R^*/R}\chi^{R_1}_a(\tau/p)\chi^{R_1}_{-\phi(a)}(\tau)~.
\eea
The second line follows by applying $S$; the minus sign is charge conjugation, $S^2=C$. Keeping the spectator module labels separate, for example by retaining its Cartan chemical potentials, determines the trace on each commutant module. Label that module by $a$ when it is paired with the spectator module $-\phi(a)$. Its symmetry-inserted trace is then $\chi^{R_1}_{-a}(p\tau)$, and its  twisted sector trace is $\chi^{R_1}_a(\tau/p)$, as used in (\ref{eq:PermutationCosetInsertion}) and (\ref{eq:PermutationCosetSpatial}). Every non-identity power of $g$ gives the same result.

\paragraph{Self-duality}
The twisted sector trace begins with $q^{-1/p}$, so $h_g=1-1/p$ and $p h_g=p-1\in\ZZ$. We may thus gauge the symmetry. The non-vacuum sectors begin at exponents $5/23$, $3/11$, and $2/5$, respectively, so they contribute no currents. For the rank exclusion used in Section \ref{sec:PermutationCosets}, the largest dimensions at ranks two and four are fourteen and fifty-two; at rank eight, $E_8$ has dimension 248, and all other semisimple possibilities have dimension at most 136. Thus an inner quantum symmetry cannot give the required ranks and dimensions. The current count and permutation argument in that section identify the resulting weight-one algebra as $R_1^{p+1}$. Its rank is twenty-four, so the lattice characterization of \cite{Dong:2002fs} and uniqueness of the corresponding Niemeier lattice identify the orbifold with $V_L$.

The surviving $R_1$ algebra has level one, so it lies in a single $R_1$ factor of the gauged theory. The Niemeier lattice symmetries permute these factors transitively, so we can identify it with the original unpermuted factor \cite{Conway:1999dov}. Any remaining action on this factor can be undone by a symmetry of the full theory. Current rotations implement inner automorphisms, while reversing all lattice momenta supplies charge conjugation for $A_2$. For $D_4^6$, order-three triality acts simultaneously on all six factors, and the order-two diagram symmetry acts simultaneously together with an exchange of two other factors. These generate all diagram symmetries of the chosen factor. We can therefore choose the identification to fix the unpermuted factor pointwise.
Since this identification fixes the spectator algebra, it also recovers the original commutant modules and their operator products, establishing self-duality of the diagonal CFT under asymmetric gauging.

\subsection{Details of the order-31 $E_{8,2}$ orbifold}
\label{app:E82Lattice}

We next give details on the calculation in Section \ref{sec:IsingEeighttwo}. As in Appendix \ref{app:exceptionalproof}, apply a modular transformation to characters with a Cartan chemical potential. For $\gamma=\left(\begin{smallmatrix}a&b\\31m&d\end{smallmatrix}\right)\in\Gamma_0(31)$, shifting the chemical potential by a lattice period gives the phase
\bea
\exp\!\left[-2\pi i\,(31m)d\left({\rho\over31}\right)^2\right]=\exp(-2\pi i\,20md)=1~.
\eea
The remaining insertion is $g^d$, which is conjugate to $g$. Thus the three inserted traces transform with the ordinary $E_{8,2}$ matrix for $\gamma$. On the other hand, the Ising characters on the right-hand side of (\ref{eq:E82CoverTraces}) are evaluated at $31\tau$, so under $\gamma$ they transform with $\rho_{\mathrm{Ising}}\!\left(\begin{smallmatrix}a&31b\\m&d\end{smallmatrix}\right)$. The Ising $T$ matrix has order 48, and its modular representation depends only on the modular transformation modulo 48. Reducing the displayed matrices modulo 48 changes $T$ to $T^{31}$ and $S$ to $\rho_{\mathrm{Ising}}(\sigma_{31})S=S$, exactly the $E_{8,2}$ modular data.

\paragraph{Self-duality} To prove self-duality, we use the diagonal embedding $E_{8,2}\subset E_{8,1}\otimes E_{8,1}$. Its commuting algebra is Ising, since the difference of central charges is $16-31/2=1/2$ \cite{Goddard:1984vk}. The branching is \cite{Forgacs:1988iw}
\bea \label{eq:EeighttwoCompletion}
V_{E_{8,1}}\otimes V_{E_{8,1}}
=\bigl(\cH^{\mathrm{Ising}}_0\otimes\cH^{E_{8,2}}_{\mathbf1}\bigr)
\oplus\bigl(\cH^{\mathrm{Ising}}_{1/2}\otimes\cH^{E_{8,2}}_{\mathbf{3875}}\bigr)\oplus\bigl(\cH^{\mathrm{Ising}}_{1/16}\otimes\cH^{E_{8,2}}_{\mathbf{248}}\bigr)~.
\eea
The rotation acts diagonally in the two level-one factors and fixes the commuting Ising algebra. In the lattice description, it is the shift $v=(\rho,\rho)/31$. The integer $31v^2/2=20$ ensures that the shifted momentum sectors admit a consistent orbifold. Gauging then replaces $E_8\oplus E_8$ by the even unimodular lattice
\bea \label{eq:EeighttwoNeighbor}
L'=\left\{(x,y)+j{(\rho,\rho)\over31}\,\,\Big| \,\,x,y\in E_8,\ j\in\ZZ,\ (x+y)\cdot\rho+20j=0\pmod{31}\right\}~.
\eea
To identify this orbifolded lattice, use coordinates in which the $E_8$ lattice consists of vectors whose components are either all integers or all half-integers, and sum to an even integer. We choose $\rho=(0,1,2,3,4,5,6,23)$, so $\rho^2=620$. The condition defining $L'$ ensures that the shift changes the squared length by an even integer. Thus every vector in $L'$ still has even squared length.

Now let $e_i$ denote the orthonormal coordinate vectors. The eight pairs in Table \ref{tabEeightNeighborRoots} define roots $\beta_i=(x_i,y_i)+(\rho,\rho)/31$ in $L'$. Their inner products give the $E_8$ Cartan matrix. The exchanged roots $\beta'_i=(y_i,x_i)+(\rho,\rho)/31$ form an orthogonal second copy. Since all inner products in $L'$ are integers and these sixteen roots have Gram determinant one, they span all of $L'$, establishing $L'\simeq E_8\oplus E_8$.

\begin{table}[t]
\centering
\begin{tabular}{c c c}
\toprule
$i$ & $x_i$ & $y_i$\\
\midrule
$1$ & $-e_1-e_8$ & $-e_6-e_8$\\
$2$ & $-e_2-e_8$ & $-e_5-e_8$\\
$3$ & $-e_3-e_8$ & $-e_4-e_8$\\
$4$ & $-e_6-e_8$ & $e_1-e_8$\\
$5$ & $-e_7-e_8$ & $-{1\over 2}\sum_{i=1}^8e_i$\\
$6$ & $0$ & $e_4-e_8$\\
$7$ & ${1\over 2}(e_1-e_2+e_3-e_4-e_5-e_6-e_7-e_8)$ & $0$\\
$8$ & $e_2-e_8$ & $-e_7-e_8$\\
\bottomrule
\end{tabular}
\caption{A simple-root system for one of the two $E_8$ factors in the order-31 orbifold lattice. Exchanging $x_i$ and $y_i$ gives the orthogonal factor.}
\label{tabEeightNeighborRoots}
\end{table}

To check that this isometry preserves the Ising algebra, let $M=\{(a,-a)\,\,|\,\,a\in E_8\}$. The intersection of $L'$ with the anti-diagonal real subspace is exactly $M$; the equation $x+y+2j\rho/31=0$ forces $j=0\pmod{31}$. Moreover, the eight vectors $x_i-y_i$ again have the $E_8$ Cartan matrix. Hence $\beta_i-\beta'_i$ span all of $M$. The two anti-diagonal lattices coincide, so the isometry preserves $M$. With a suitable choice of phases on momentum states, it also preserves the Ising stress tensor constructed from $M$ \cite{Chen:2014aia}. Taking the commutant of this Ising algebra then recovers $E_{8,2}$.

\subsection{Details of the order-47 Baby Monster orbifold}
\label{app:BabyScalar}

Finally, we derive the order-47 trace identities (\ref{eq:BabyCoverTraces}) used in Section \ref{sec:IsingBaby}, using the Monster branching (\ref{eq:BabyCompletion}). Let $t$ be the central involution in $2.\mathrm{BM}$. It lies in Monster class $2A$ and acts by $+1,+1,-1$ on the three branches. The element $g$ lies in class $47A$ or $47B$, while $tg$ lies in class $94A$ or $94B$; each pair has the same McKay-Thompson series.

Write the three Baby Monster symmetry-inserted traces as
\bea
F^{\mathrm{BM}}_h(\tau)=\Tr_{\cH^{\mathrm{BM}}_h}\!\left(gq^{L_0-47/48}\right)~,
\hspace{0.4 in}h=0,3/2,31/16~.
\eea
Tracing $g$ and $tg$ on (\ref{eq:BabyCompletion}), and tracing $g$ on the $t$-twisted Monster module, gives
\bea \label{eq:BabyMoonshineSystem}
\chi^{\mathrm{Ising}}_0F^{\mathrm{BM}}_0
+\chi^{\mathrm{Ising}}_{1/2}F^{\mathrm{BM}}_{3/2}
+\chi^{\mathrm{Ising}}_{1/16}F^{\mathrm{BM}}_{31/16}
&=&T_{47A}(\tau)~,\no\\
\chi^{\mathrm{Ising}}_0F^{\mathrm{BM}}_0
+\chi^{\mathrm{Ising}}_{1/2}F^{\mathrm{BM}}_{3/2}
-\chi^{\mathrm{Ising}}_{1/16}F^{\mathrm{BM}}_{31/16}
&=&T_{94A}(\tau)~,\no\\
\chi^{\mathrm{Ising}}_{1/2}F^{\mathrm{BM}}_0
+\chi^{\mathrm{Ising}}_0F^{\mathrm{BM}}_{3/2}
+\chi^{\mathrm{Ising}}_{1/16}F^{\mathrm{BM}}_{31/16}
&=&T_{94A}(\tau/2)~.
\eea
All factors on the left are evaluated at $\tau$, and $T_{47A}$ and $T_{94A}$ denote the corresponding McKay-Thompson series. The first two lines follow directly from the untwisted Monster decomposition. We now justify the last line.

To do so, put $x=tg$, so that $x$ has order 94, $t=x^{47}$, and $g=x^{48}$. The modular transformation $\left(\begin{smallmatrix}1&1\\47&48\end{smallmatrix}\right)$ takes an untwisted insertion of $x$ to the boundary conditions $(t,g)$. Modular covariance therefore gives $T_{94A}((\tau+1)/(47\tau+48))$ up to a constant \cite{Carnahan:2012gx}. The Atkin-Lehner symmetry $\left(\begin{smallmatrix}2&1\\94&48\end{smallmatrix}\right)$ of $T_{94A}$ sends $\tau/2$ to this argument \cite{Conway:1979qga}, so
\bea
\Tr_{V^\natural(t)}gq^{L_0-1}=T_{94A}(\tau/2)~,
\eea
as claimed above.

To identify the functions $F^{\mathrm{BM}}_h$, we now compare these results with bilinears in Ising characters at $\tau$ and $47\tau$. Recall that the Ising conductor is 48. Since $47=-1\pmod{48}$, the functions at $\tau$ and $47\tau$ have opposite $T$ phases. Their diagonal pairing is invariant under $\Gamma_0(47)$. It is also invariant under the Fricke transformation, 
\bea
W_{47}:\,\,\,\tau\longmapsto-{1\over47\tau}~.
\eea
This sends the pair of arguments $(\tau,47\tau)$ to $(-1/47\tau,-1/\tau)$. Applying the Ising $S$ matrix to both characters then exchanges their arguments, and orthogonality of $S_{\mathrm{Ising}}$ leaves the diagonal pairing unchanged. Its vacuum term is $q^{-(1+47)/48}=q^{-1}$. These facts suggest the scalar identity
\bea 
\label{eq:BabyScalarFirst}
T_{47A}(\tau)=\chi^{\mathrm{Ising}}_0(\tau)\chi^{\mathrm{Ising}}_0(47\tau)+\chi^{\mathrm{Ising}}_{1/2}(\tau)\chi^{\mathrm{Ising}}_{1/2}(47\tau)+\chi^{\mathrm{Ising}}_{1/16}(\tau)\chi^{\mathrm{Ising}}_{1/16}(47\tau)~.
\eea
The group generated by $\Gamma_0(47)$ and $W_{47}$ is called the Fricke group. The vacuum pole is the only pole modulo this group, and the constant term vanishes. These conditions then uniquely determine the McKay-Thompson series, proving the identity.

Next define
\bea
\mathcal Z_1(\tau)&:=&\chi^{\mathrm{Ising}}_0(\tau)\chi^{\mathrm{Ising}}_0(47\tau)+\chi^{\mathrm{Ising}}_{1/2}(\tau)\chi^{\mathrm{Ising}}_{1/2}(47\tau)-\chi^{\mathrm{Ising}}_{1/16}(\tau)\chi^{\mathrm{Ising}}_{1/16}(47\tau)~,\no\\
\mathcal Z_2(\tau)&:=&\chi^{\mathrm{Ising}}_{1/2}(\tau)\chi^{\mathrm{Ising}}_0(47\tau)+\chi^{\mathrm{Ising}}_0(\tau)\chi^{\mathrm{Ising}}_{1/2}(47\tau)+\chi^{\mathrm{Ising}}_{1/16}(\tau)\chi^{\mathrm{Ising}}_{1/16}(47\tau)~.
\eea
We will identify these with $T_{94A}(\tau)$ and $T_{94A}(\tau/2)$, respectively. To do so, we first show that $\mathcal Z_1(\tau)=\mathcal Z_2(2\tau)$ by comparing their modular transformations and their singular terms.

In the ordering $(0,1/2,1/16)$, the matrices
\bea
D=\mathrm{diag}(1,1,-1)~,\hspace{0.4 in}
P=S_{\mathrm{Ising}}DS_{\mathrm{Ising}}=
\begin{pmatrix}
0&1&0\\
1&0&0\\
0&0&1
\end{pmatrix}
\eea
record the character pairings in $\mathcal Z_1$ and $\mathcal Z_2$. The matrix $D$ supplies the minus sign in $\mathcal Z_1$, while $P$ exchanges the vacuum and weight-$1/2$ characters.
For a transformation
\bea
\gamma=\begin{pmatrix}a&b\\94m&d\end{pmatrix}\in\Gamma_0(94)~,
\eea
the characters at $\tau$ and $47\tau$ transform using the matrices
\bea
\begin{pmatrix}a&b\\94m&d\end{pmatrix}~,\hspace{0.4 in}
\begin{pmatrix}a&47b\\2m&d\end{pmatrix}~,
\eea
respectively. The Ising modular representation depends only on these matrices modulo 48. Since $47=-1\pmod{48}$, the two Ising modular matrices are complex conjugates. Moreover, $D$ commutes with $T$ and $ST^2S$, which generate $\Gamma_0(2)$, and hence with the matrices for its subgroup $\Gamma_0(94)$. Unitarity then implies that the pairing with $D$ is invariant, so $\mathcal Z_1$ is a modular function for $\Gamma_0(94)$.
Applying $S$ gives
\bea
\mathcal Z_1\!\left(-{1\over94\tau}\right)
&=&\chi^{\mathrm{Ising}}_0(94\tau)\chi^{\mathrm{Ising}}_{1/2}(2\tau)+\chi^{\mathrm{Ising}}_{1/2}(94\tau)\chi^{\mathrm{Ising}}_0(2\tau)+\chi^{\mathrm{Ising}}_{1/16}(94\tau)\chi^{\mathrm{Ising}}_{1/16}(2\tau)\no\\
&=&\mathcal Z_2(2\tau)~,
\eea
where we used $S_{\mathrm{Ising}}DS_{\mathrm{Ising}}=P$. This relates the two functions by the Fricke transformation. Since the Fricke transformation maps $\Gamma_0(94)$ to itself by conjugation, it also shows that $\mathcal Z_2(2\tau)$ is a modular function for $\Gamma_0(94)$.

Now note that both $\mathcal Z_1(\tau)$ and $\mathcal Z_2(2\tau)$ are holomorphic in the upper half-plane, with their only possible poles occuring at the four cusps $\infty$, $0$, $1/2$, and $1/47$. Applying the Ising matrices for $\mathds{1}$, $S$, $\left(\begin{smallmatrix}1&0\\2&1\end{smallmatrix}\right)$, and $\left(\begin{smallmatrix}1&0\\47&1\end{smallmatrix}\right)$ shows that their polar parts agree and are, respectively,
\bea
q^{-1}~,
\hspace{0.4 in}q^{-1/94}~,
\hspace{0.4 in}e^{2\pi i\,23/47}q^{-1/47}~,
\hspace{0.4 in}-q^{-1/2}~.
\eea
Since all negative-power terms agree, the difference $\mathcal Z_1(\tau)-\mathcal Z_2(2\tau)$ has no poles. The maximum principle therefore implies that it is constant \cite{DHoker:2024xor}. Both series have vanishing constant terms at infinity, so this constant is zero, proving $\mathcal Z_1(\tau)=\mathcal Z_2(2\tau)$.

To further identify these with the relevant McKay-Thompson series, note that the same Ising $S$ and $T$ matrices give
\bea
\cZ_{1}\left({2\tau-1\over94\tau-46}\right)=\cZ_{2}(2\tau)=\cZ_{1}(\tau)~.
\eea
This is an additional symmetry $W_2$ at level 94. Together with $\Gamma_0(94)$ and the Fricke symmetry already established, it generates $\Gamma_0(94)^+$. All cusps are equivalent under this group, so the single normalized pole $q^{-1}$ and zero constant term identify $\cZ_{1}$ as $T_{94A}=T_{94B}$ \cite{Conway:1979qga}, from which we also identify $\cZ_2$ as  $T_{94A}(\tau/2)$. 
We may finally use (\ref{eq:BabyMoonshineSystem}), (\ref{eq:BabyScalarFirst}), and the two identities just proved to fix the functions $F^{\mathrm{BM}}_h$, which gives the three traces in (\ref{eq:BabyCoverTraces}).

\bibliographystyle{ytamsalpha}
\bibliography{references}
\end{document}